\documentclass{article}

\usepackage[aux]{rerunfilecheck}
\usepackage{hyperref}
\usepackage{amsmath}
\usepackage{amssymb}
\usepackage{mathtools}
\usepackage[numbers,sort&compress]{natbib}
\usepackage[autostyle]{csquotes}
\usepackage{xfrac}
\usepackage{float}
\usepackage{graphicx}
\usepackage{color}
\usepackage{bbm} 
\usepackage{placeins}
\usepackage{booktabs}
\usepackage{subcaption} 
\usepackage{geometry}

\newcommand{\dNeff}{\ensuremath{\Delta N_{\mathrm{eff}}}}

\title{\bf Light Sterile Neutrinos in the Early Universe: Effects of Altered Dispersion Relations and a coupling to Axion-Like Dark Matter}

\author{Dominik Hellmann$^1$, Heinrich P\"as$^1$
\smallskip
\\
{\it $^1$ Fakult\"at f\"ur Physik,
Technische Universit\"at Dortmund,
Germany}
}

\begin{document}

\maketitle

\begin{abstract}
We investigate the cosmological consequences of light sterile neutrinos with altered dispersion relations (ADRs) 
and couplings to an ultra-light, axion-like scalar field. 
In particular we study the impact on the number of additional, light, fermionic degrees of freedom and primordial nucleosynthesis.
While the ADR leads to a new potential term in the Hamiltonian, 
the coupling to the scalar field results in a time dependent, effective mass contribution. 
We solve the quantum kinetic equations (QKEs) for the neutrino density matrix 
and find that in certain parameter regions both new physics effects can individually yield a suppressed population of sterile neutrino species 
and the correct observed amount of helium in nucleosynthesis.
Combining both effects opens up new patches of parameter space excluded by experimental bounds applying to models featuring only one of the effects.

\end{abstract}

\section{Introduction}
\label{sec:intro}
There exist several anomalies in short baseline (SBL) neutrino experiments that point towards one or more sterile neutrinos in the 
\(\mathcal{O}(1\,\mathrm{eV})\) range ~\cite{MiniBooNE:2018esg, LSND:2001aii, Mention:2011rk, Giunti:2010zu}. 
Even after the recently published MicroBooNE data~\cite{MicroBooNE:2021tya, MicroBooNE:2021pvo, TheMicroBooNECollaboration:2021cjf, MicroBooNE:2021nxr} had disfavored the excess observed in the MiniBooNE experiment, 
sterile neutrinos in this mass range remain an interesting option for Standard Model (SM) extensions~\cite{Denton:2021czb,Acero:2022wqg}.

As has been shown many years ago, the simple solution of just adding a few sterile neutrino generations in the correct mass range to the SM is not viable, since it solves the problem only for SBL experiments while long baseline (LBL) and atmospheric neutrino oscillation experiments do not observe any anomalies~\cite{MINOS:2017cae, T2K:2019efw, NOvA:2017geg, NOvA:2021smv, Super-Kamiokande:2014ndf, IceCube:2016rnb} at the same baseline-energy ratio  \(L / E\).

Therefore, if the SBL anomalies are indeed explained by light sterile neutrinos, the oscillation probability cannot depend on \(L / E\) in the same way as it would be the case in the standard description.
There exist several proposals to resolve this tension (for a recent review see~\cite{Acero:2022wqg}). One such possibility is to include an additional potential term for sterile neutrinos in the Hamiltonian that parametrizes possible new physics effects.
As a consequence, an energy dependent mixing pattern arises between active and sterile neutrinos.
In this work, we mainly concentrate on a model proposed and investigated in refs.~\cite{Pas:2005rb, Marfatia:2011bw, Aeikens:2014yga, Aeikens:2016rep, Doring:2018cob, Jang:2018rkc, Barenboim:2019hso} where the active-sterile mixing becomes maximal at some resonance energy \(E_{\mathrm{res}}\) and nearly vanishes at higher energies.
Such altered dispersion relations (ADRs) for example arise if sterile neutrinos can take shortcuts through asymmetrically warped extra dimensions that are unaccessible to particles charged under the SM gauge group~\cite{Doring:2018ncz}.
In order to solve the SBL anomalies, one has to require that \(E_{\mathrm{res}}\) is located within the energy ranges probed by these experiments such that oscillations incorporating the new heavier mass eigenstate associated with the sterile neutrino can reproduce the observed pattern.
At higher energies, the sterile neutrinos decouple from the active ones due to mixing suppression and the standard oscillation pattern is restored in the
energy ranges typically probed by LBL experiments.

While this modification may help to resolve the 
tensions encountered between SBL and LBL experiments, there remains a severe tension with cosmological observations.
For example, the effective number of neutrino generations \(N_{\mathrm{eff}} := 3 + \Delta N_{\mathrm{eff}}\) has been shown to be very close to 3 at the time of recombination according to recent Planck data~\cite{Planck:2018vyg}, while it could be altered dramatically if new, light, sterile degrees of freedom are equilibrated in the early universe.
Furthermore, the helium fraction produced in big bang nucleosynthesis (BBN), \(Y_{\mathrm{^4He}}\), is sensitive to two effects.
The first and most significant one is the faster expansion of the universe caused by the presence of additional light degrees of freedom leading to an earlier freeze out of neutron-proton reactions.
The second effect is the possible conversion of electron neutrinos to sterile neutrinos also resulting in a potentially earlier freeze out since electron neutrinos are a substantial part of almost all reactions keeping neutrons and protons in equilibrium with the plasma.
Sterile neutrinos with ADRs may also ameliorate this cosmological tension by suppressing the population of the sterile flavor at high energies in the early universe, as has been suggested in~\cite{Aeikens:2016rep}.

Another idea that has been proposed to reconcile light sterile neutrinos with cosmological data employs an axion-like particle (ALP) coupling to the sterile neutrinos via a Yukawa interaction term~\cite{Berlin:2016woy, Zhao:2017wmo, Farzan:2019yvo, Cline:2019seo, Huang:2021kam, Huang:2022wmz} (for non-cosmological applications of this particular class of models see also~\cite{Krnjaic:2017zlz, Brdar:2017kbt, Liao:2018byh, Capozzi:2018bps, Reynoso:2022vrn}).
The additional, ultra-light scalar field \(\phi\) is assumed to be an approximately homogeneous condensate behaving like a classical field.
Its time evolution is governed by the Klein-Gordon equation in an expanding universe and influences the effective mass of the sterile neutrino and hence the mixing of the sterile and active neutrino.

Note that similar ideas pursued in the literature~\cite{Hannestad:2013ana, Dasgupta:2013zpn, Mirizzi:2014ama, Tang:2014yla, Forastieri:2017oma, Archidiacono:2020yey,Davoudiasl:2023uiq} also involve the coupling of sterile neutrinos to dark scalars or massive gauge bosons inducing an effective potential in the oscillation Hamiltonian suppressing the thermalization of sterile degrees of freedom in the early universe.
In the models considered in this work, the \(\nu_s\)-\(\phi\) coupling is small and consequently, the induced matter potential and non-linear self interactions of \(\nu_s\) are strongly suppressed.
The additional sterile neutrino mass still remains relevant at early times due to the large initial amplitude of the classical part of \(\phi\).
Furthermore, the ADR potential included in the oscillation Hamiltonian is not assumed to be induced by interactions as described above and can be sufficiently weakly temperature dependent such that this dependence can be neglected.

In this paper we discuss both, the individual cosmological effects of and the interplay between altered dispersion relations and a time varying sterile mass induced by a light scalar field.
In order to estimate the influence of the full model on the chosen set of cosmological quantities, i.e. the effective number of additional neutrino generations \(\Delta N_{\mathrm{eff}}\) and the helium fraction \(Y_{\mathrm{^4He}}\), we solve the quantum kinetic equations (QKEs) for the neutrino density matrix \(\varrho\).
This defines the main task of this work, since from \(\varrho\) we can derive the neutrino phase space distributions and energy densities.
In order to simplify our framework for this study, we only consider a two flavor system of one active and one sterile flavor.
Moreover, we take the active neutrino to be the electron neutrino to be able to draw realistic conclusions about nucleosynthesis.

This paper is organized as follows: in section~\ref{sec:model}, we describe the model in detail and define the parameter space and crucial parameters.
Section~\ref{sec:density_matrix} is concerned with setting up and numerically solving the Boltzmann equations.
Furthermore, we discuss the proper definition of the density matrix.
In section~\ref{sec:results}, we present results for the final helium abundances and \(\Delta N_{\mathrm{eff}}\) for various benchmark points in the model parameter space.
Finally in section~\ref{sec:conclusions}, we draw our conclusions.

\section{Sterile Neutrinos with Altered Dispersion Relations coupling to Axion-Like Dark Matter}
\label{sec:model}

\subsection{Sterile Neutrinos with Altered Dispersion Relations}
\label{ssec:model_adr}
Assuming that the relativistic dispersion relations of neutrinos are altered by some unspecified new physics effect such as the presence of asymmetrically warped extra dimensions\footnote{The asymmetric warping leads to the effect that sterile neutrinos propagating through the extra dimension between two points \(x_1\) and \(x_2\) on the SM brane need less time to complete their journey than SM neutrinos traveling between those points on geodesics bound to the SM brane~\cite{Pas:2005rb,Doring:2018ncz}.
An observer located on the brane will hence come to the conclusion that the usual energy momentum relation does not hold for sterile neutrinos and needs to incorporate an effective potential into \(E^2 = p^2 + m^2\) in order to describe their behavior using brane bound quantities.} leads to new terms in the propagation Hamiltonian.
Employing the usual ultra-relativistic expansion of the neutrino dispersion relation yields
\begin{align}
    \mathcal{H}(p) &= \frac{1}{2 p} M^{\dagger} M + V_{s}(p)\,,\\
    V_{s}(p) &= -\frac{bp}{2} \mathbb{P}_{s}\,,
\end{align}
where \(p\) is the average neutrino momentum, \(b\) is the so called ADR parameter controlling the strength of the ADR effect, \(\mathbb{P}_{s}\) is the sterile neutrino projector and \(M\) is the neutrino mass matrix.

In the flavor basis, the last two quantities read 
\begin{align}
    M &= \begin{pmatrix}
        m_{ee} & m_{es}\\
        m_{es}^\ast & m_{ss}\\
    \end{pmatrix}\,, \\
    \mathbb{P}_{s} &= \begin{pmatrix}
        0 & 0\\
        0 & 1\\
    \end{pmatrix}\,,
\end{align}
for the \(2 \times 2\) neutrino system under consideration.
Furthermore, we choose all mass parameters to be real valued meaning that there is no CP violation in the active-sterile mixing.
To be specific, we use
\begin{align}
    m_{ee} \approx 0\,\mathrm{eV}\,, \quad m_{es} \approx 0.12\,\mathrm{eV}\,, \quad m_{ss} \approx 1.1\,\mathrm{eV}\,,
\end{align}
in accordance with ref.~\cite{Cline:2019seo} and fits to SBL data.

Since we consider the system in the early universe, we also have to include a potential for the electron neutrino induced by elastic scattering processes modifying the effective masses of the neutrino matter eigenstates.
Therefore, the full Hamiltonian reads~\cite{Sigl:1993ctk}
\begin{align}
    \mathcal{H}(p) &= \frac{1}{2 p} M^{\dagger} M + V_{e}(p) + V_{s}(p) \label{eq:Ham}\\
    V_{e}(p) &= -\frac{8 \sqrt{2} G_F p}{3} \left(\frac{\rho_{e^{\pm}}}{m_W^2} + \frac{\rho_{\nu_e}}{m_Z^2}\right) \mathbb{P}_{e}\,,
\end{align}
with the Fermi coupling \(G_F\), the electron and neutrino energy densities \(\rho_{\alpha}\) and the \(W,Z\)-Boson masses \(m_{W,Z}\).
Here, we neglected the usual MSW potential proportional to the particle-antiparticle asymmetries because we assume them to be small (of the order of the baryon asymmetry).
Thus, \(V_{e}\) only contains the more significant higher order elastic scattering contributions from the inverse mass expansion of the \(W,Z\)-Boson propagators.

In order to analyze the resonance structure of this two flavor system, we diagonalize the Hamiltonian with the general ansatz
\begin{align}
    U(\theta(p)) := \begin{pmatrix}
        \cos(\theta(p)) & \sin(\theta(p)) \\
        -\sin(\theta(p)) & \cos(\theta(p)) \\
    \end{pmatrix} \label{eq:mixmat}
\end{align}
and find the following relation for the mixing angle
\begin{align}
    \tan(2\theta(p)) &= \frac{2 m_{es} (m_{ee} + m_{ss})}{(m_{ss}^2 - m_{ee}^2) + 2p (V_s(p) - V_e(p))} \label{eq:tan2th}\,.
\end{align}
The mixing becomes maximal as soon as \(\theta = \pi / 4\) because then electron and sterile neutrinos equally constitute both mass eigenstates and the mass gap between these two eigenstates is minimal.
This in turn leads to higher transition rates between active and sterile neutrinos in the energy regime close to the resonance.

From eq.~\eqref{eq:tan2th} we can infer the resonance condition
\begin{align}
    \theta \rightarrow \frac{\pi}{4} &\Rightarrow \tan(2\theta(p)) \rightarrow \pm \infty \nonumber\\
    &\Rightarrow (m_{ss}^2 - m_{ee}^2) + 2p (V_s(p) - V_e(p)) \rightarrow 0^{\pm} \label{eq:res_cond}\,.
\end{align}
The momentum \(p_{\mathrm{res}}\) fulfilling the condition~\eqref{eq:res_cond} is called the resonance momentum.
As follows from eq.~\eqref{eq:res_cond}, the interplay of the two appearing potentials and the neutrino masses determines the resonance structure of the system.

This will become important again as soon as we study the effects of resonant conversion between sterile and active neutrinos in combination with collisions in the early universe plasma.

\subsection{Sterile Neutrinos coupling to Axion-Like Dark Matter}
\label{ssec:model_alp}
In addition to the ADR effects, we also introduce an ultra-light, real, scalar field.
The corresponding scalar particles are assumed to be produced non-thermally and to form a coherent condensate.
This condensate behaves like a homogeneous, classical field in an expanding universe and its time evolution is given by~\cite{Cline:2019seo}
\begin{align}
    \phi(t) &= \phi_0 \eta(t)\\
    \eta(t) &= 1.08 \frac{J_{\frac{1}{4}}(m_{\phi} t)}{\sqrt[4]{m_{\phi} t}} \quad \mathrm{with} \quad \lim\limits_{t \rightarrow 0} \eta(t) = 1\,. \label{eq:phi_hat}
\end{align}
Here, \(J_{\frac{1}{4}}\) is the regular Bessel \(J\) function of fractional order \(1 / 4\), \(m_\phi\) is the mass of the scalar and \(t\) is the cosmic time.
Moreover, we allow for a feeble\footnote{The coupling has to be feeble in order for the scalar field not to thermalize if the sterile neutrinos are thermalized via oscillations.} coupling to sterile neutrinos via a Yukawa interaction term in its Lagrangian,
\begin{align}
    \mathcal{L}(\phi, \partial_{\mu} \phi) &= \frac{1}{2}(\partial_{\mu} \phi) (\partial^{\mu} \phi) - \frac{m_{\phi}^{2}}{2} \phi^{2} - \frac{\lambda}{2} \phi \bar{\nu}_{s} \nu_{s} \,.
\end{align}
Since we assume the coupling constant \(\lambda\) to be very small, we neglect possible interactions between \(\nu_s\) and \(\phi\) quanta and interactions of \(\nu_s\) with itself mediated by \(\phi\). 
Nevertheless, we account for an additional time dependent sterile neutrino mass term
\begin{align}
    m_{\mathrm{eff}}(t) &= m_{ss} + \lambda \phi_{0} \eta(t) \,,
\end{align}
modifying the mixing between electron and sterile neutrinos.
Although this additional mass term is proportional to \(\lambda\), it is indeed significant if the amplitude \(\phi_0\) of the classical component of \(\phi\) is sufficiently large.
Assuming \(\phi\) to contribute a substantial amount to the dark matter density, one can deduce~\cite{Cline:2019seo}
\begin{align}
    \phi_0 \sim 10^{15} \, \mathrm{GeV} \left(\frac{10^{-15}\, \mathrm{eV}}{m_\phi}\right)^{\frac{1}{4}} \,.
\end{align} 
Thus, for scalar masses on the order of \(m_{\phi} \sim 10^{-10}\,\mathrm{eV}\) one can still obtain \(\lambda \cdot \phi_0 \sim 100\,\mathrm{eV}\) for \(\lambda \sim 10^{-21}\).

At early times, the mass contribution remains approximately constant leading to a constant sterile neutrino mass \(m_{\mathrm{eff}} \approx m_{ss} + \lambda \phi_0\).
If \(\phi_{0}\) is sufficiently large, the effective mixing of electron and sterile neutrinos is negligible since \(\tan(2\theta)\) is suppressed by the large mass gap \(m_{\mathrm{eff}}^2 - m_{ee}^2\) in the denominator of eq.~\eqref{eq:tan2th} with \(m_ss\) replaced by \(m_{\mathrm{eff}}\).
Hence, sterile neutrinos are not populated via oscillations at these early times.

From the point in time where the Bessel function starts its damped oscillating behavior (\(t \sim 1 / m_{\phi}\)) the effective sterile neutrino mass approaches \(m_{ss}\), allowing for significant oscillations again.
Therefore, the neutrino oscillation behavior depends on the mass parameter \(m_{\phi}\), i.e. the smaller \(m_{\phi}\) the longer \(\phi(t)\) will remain approximately constant and active-sterile mixing will be suppressed by \(m_{s0}\).

\subsection{Central Quantities and Parameter Space}
\label{ssec:model_obs}
In the following analysis, we focus on two major quantities:
\begin{enumerate}
    \item The effective number of additional neutrino generations
    \begin{align}
        \Delta N_{\mathrm{eff}}(t) :&= \frac{8}{7} \left(\frac{11}{4}\right)^{\frac{4}{3}} \sum_{k \in \mathcal{V}} \frac{\rho_k(t)}{\rho_{\gamma}(t)} - 3\,, \quad \mathrm{with} \quad \mathcal{V} = \{e, \mu, \tau, s\}\,.
    \end{align}
    Here, \(\rho_k(t)\) are the neutrino energy densities at time \(t\) and \(\rho_\gamma(t)\) is the corresponding photon density\footnote{By using the energy density in the definition of \(\dNeff\) we slightly differ from the methods employed in ref.~\cite{Cline:2019seo} where the number density is used instead.}.
    Moreover, we need to incorporate factors of \(8 / 7\) and \((11 / 4)^{4 / 3}\) to directly compare fermionic and bosonic energy densities with a temperature deviation of \((11 / 4)^{1 / 3}\) to each other.
    \item The helium mass fraction
    \begin{align}
        Y_{^{4}\mathrm{He}} &:= \frac{4 n_{\mathrm{He}}}{n_{\mathrm{B}}}\,,
    \end{align}
    with the helium and baryon number densities \(n_{\mathrm{He}}\), \(n_{\mathrm{B}}\).
\end{enumerate}
In order to compare these quantities with observations, these observables have to be computed at different times in the cosmic evolution.
The value of the number of additional neutrino generations \(\Delta N_{\mathrm{eff}}(t)\) is inferred from the Hubble rate measurement from the CMB~\cite{Planck:2018vyg} and thus needs to be known at the time of the last photon scattering, while the value of the helium fraction \(Y_{^{4}\mathrm{He}}\) has to be computed shortly after BBN.
It is, however, sufficient to evaluate \(\dNeff\) directly after \(e^\pm\) annihilation since from there on it remains constant.
This is because after \(e^\pm\) annihilation the total neutrino energy density only changes due to the expansion of the universe\footnote{We prove that the total neutrino energy density only changes due to the expansion of the universe after neutrino decoupling in Appendix~\ref{app:rho_nu} using the neutrino density matrix formalism introduced in the next section.} and by normalizing it to the photon energy density we cancel the dependence on the scale factor.

The helium fraction remains constant right after BBN, hence it is appropriate to evaluate \(Y_{4^\mathrm{He}}\) as soon as deuterium dissociation has ceased to be efficient.
Hence, our analysis solely focuses on the era of radiation domination.

Finally, we want to discuss the three dimensional parameter space of the model under consideration.
It is parameterized by
\begin{enumerate}
    \item The scalar field mass \(m_{\phi}\) determining when the scalar field starts to oscillate.
    \item The amplitude \(m_{s0} := \lambda \phi_0\) of the additional mass contribution for the sterile neutrino.
    \item The ADR parameter \(b\).
\end{enumerate}
In the following, we assume a range for \(m_{s0} \in [10\,\mathrm{eV}, 250\,\mathrm{eV}]\) in accordance with ref.~\cite{Cline:2019seo}.
Furthermore, for \(m_{\phi}\) we choose the interval of possible values to be \([10^{-22}\,\mathrm{eV}, 10^{-14}\,\mathrm{eV}]\) since for \(m_{\phi} \leq  10^{-22}\,\mathrm{eV}\) the scalar field starts oscillating so late that the effective sterile neutrino mass remains constant during the considered temperatures.
Thus, for \(m_{\phi} \lesssim 10^{-22}\,\mathrm{eV}\) our scenario becomes independent of \(m_{\phi}\) and yields the same constraints on \(m_{s0}\) as for \(m_{\phi} \sim 10^{-22}\,\mathrm{eV}\).

At values larger than \(m_{\phi} = 10^{-14}\,\mathrm{eV}\) the addition of the scalar field to the model becomes meaningless since the sterile neutrino mass already gets close to its \textit{bare} value at the relevant temperatures and the sterile species equilibrates.

For the ADR parameter, we choose benchmark values in \(I_{b} = [0, \infty)\), i.e. we consider anything between zero ADR potential and an arbitrarily large ADR effect.
A special point in this parameter range is \(b \sim 10^{-17}\), since this is the order of magnitude needed in order to explain SBL anomalies~\cite{Doring:2018cob}.
The reason why we allow for arbitrarily high (low) ADR parameters in the early universe is that there might be some mechanism in the extradimensional realization of ADRs causing the curvature of the extra dimension to change from early times until today and correspondingly implying the ADR parameter to decrease (increase), accordingly.

\section{Neutrino Quantum Kinetic Equations and Numerical Strategy}
\label{sec:density_matrix}
The density matrix describing the oscillations of neutrinos in the early universe is defined as the thermal average of creation and annihilation operators, \(a_{j}^{(\dagger)}(\vec{p})\) of neutrino mass eigenstates, i.e.~\cite{Sigl:1993ctk,Vlasenko:2013fja}
\begin{align}
    (2 \pi)^3 \delta^{3}(\vec{p} - \vec{q}) \varrho_{jk}(\vec{p}) := \langle a_k(\vec{p})^{\dagger} a_j(\vec{q}) \rangle
\end{align}
where the thermal average of an operator \(\hat{O}\) is defined using the density operator\footnote{Despite their very similar names the density operator and density matrix are by no means equivalent quantities.} \(\hat{\Pi}\) of the thermal system as usual \(\langle \hat{O} \rangle := \mathrm{Tr}(\hat{\Pi} \hat{O})\).

Thus, if we consider the diagonal elements (\(j = k\)) of the density matrix it simplifies to the thermal average of the occupation number operator of \(\nu_{k}\) which in turn results in its phase space distribution function.
This already gives some intuition of its physical meaning:
The diagonal of the density matrix contains the information how many \(\nu_k\) momentum states on average are occupied in the system.
Moreover, by inspecting the off-diagonal elements of \(\varrho\), we obtain information about the average correlation between \(\nu_{j}(p)\) and \(\nu_k(p)\).
Such correlations arise for example due to neutrino oscillations.
This makes the density matrix the quantity of choice if we want to consider incoherent particle collisions and oscillations in a thermal environment~\cite{Stodolsky:1986dx} at the same time.

In the following, we will mainly work in the flavor basis instead of the mass basis because the collision terms and Hamiltonian potentials are easier to calculate in the flavor basis.
Therefore, we need to transform \(\varrho_{jk}\) into this basis using the neutrino mixing matrix \(U\) from eq.~\eqref{eq:mixmat} via
\begin{align}
    \varrho_{jk}^{f} = \sum_{l,m = 1}^{2} U_{jl} \varrho_{lm} (U^{\dagger})_{mk}\,.
\end{align}
From now on, we will work with \(\varrho^{f}\) only and drop the superscript \(f\).

\subsection{Quantum Kinetic Equations for the Density Matrix}
\label{ssec:Boltzmann}
The time evolution of \(\varrho\) in an expanding, homogeneous and isotropic universe is governed by a Boltzmann-like, quantum kinetic equation (QKE)~\cite{Sigl:1993ctk,Vlasenko:2013fja,deSalas:2016ztq,Blaschke:2016xxt,Gariazzo:2019gyi}
\begin{align}
    (\partial_t - p H \partial_p) \varrho(t, p) &= -i[\mathcal{H}(t, p), \varrho(t, p)] + \mathcal{C}[t, p, \varrho]\,, \label{eq:EoM_}
\end{align}
where \(p\) is the modulus of the neutrino momentum, \(t\) is the cosmic time, \(H\) is the Hubble rate, \(\mathcal{H}\) is the Hamiltonian from eq.~\eqref{eq:Ham} and \(\mathcal{C}\) is the collison operator.

The convectional derivative operator, \(\partial_t - p H \partial_p\), on the left hand side includes the effect of the expansion of the universe redshifting the neutrino momentum as \(p \propto a^{-1}\), where \(a\) is the scale factor of the Robertson Walker metric.
Furthermore, the right hand side contains the commutator of the neutrino Hamiltonian with the density matrix and the collision operator \(\mathcal{C}\).
While the commutator part governs the evolution of \(\rho\) due to neutrino oscillations, the collision part determines how many neutrinos are annihilated, created or scattered to other momentum modes in interactions with the background plasma.
Since there are also neutrinos in the background plasma, this last term is non-linear.

The collision operator is the sum of the individual collision operators corresponding to a scattering process involving neutrinos
\begin{align}
    \mathcal{C}[t, p, \varrho] &= \sum_{k \in \mathcal{P}} \mathcal{C}_{k}[t, p, \varrho]\,,
\end{align}
where the set of all processes \(\mathcal{P}\) contains the interactions given in table~\ref{tab:processes}.
Here we neglect neutrino-nucleon scattering processes due to strong Boltzmann suppression of the nucleon distribution functions at temperatures of \(\mathcal{O}(100\,\mathrm{MeV})\) and smaller.
\begin{table}
    \centering
    \caption{All relevant processes considered in the neutrino collision terms.}
    \label{tab:processes}
    \begin{tabular}{cc}
        \toprule
        k & process \\
        \midrule
        1 & \(\nu e^{-} \leftrightarrow \nu e^{-}\)\\
        2 & \(\nu e^{+} \leftrightarrow \nu e^{+}\)\\
        3 & \(\nu \bar{\nu} \leftrightarrow e^{-} e^{+}\)\\
        4 & \(\nu \nu \leftrightarrow \nu \nu\)\\
        5 & \(\nu \bar{\nu} \leftrightarrow \nu \bar{\nu}\)\\
        \bottomrule
    \end{tabular}
\end{table}

For example, the collision term for the process \(\nu e^{-} \leftrightarrow \nu e^{-}\) is given by
\begin{align}
    \mathcal{C}_{1}[t, p, \varrho] = &\frac{8 G_F^2}{p} \int \mathrm{d}^3\vec{\pi}_1 \, \mathrm{d}^3\vec{\pi}_2 \, \mathrm{d}^3\vec{\pi}_3\,
    (2\pi)^4 \delta^{4}(p^{\mu} + p_{1}^{\mu} - p_{2}^{\mu} - p_{3}^{\mu}) \nonumber \\
    &\times \left\{ 4 g_{l}^{2} (p_{\alpha}p_{3}^{\alpha}) (p_{1\beta}p_{2}^{\beta})
    + 4 g_{R}^{2} (p_{\alpha}p_{1}^{\alpha}) (p_{2\beta}p_{3}^{\beta})
    - 4 g_{L}g_{R} (p_{\alpha}p_{2}^{\alpha}) m_e^2\right\} \nonumber \\
    &\times \phi_{\nu e^{-} \nu e^{-}}(t, p^{\mu},p_{1}^{\mu},p_{2}^{\mu},p_{3}^{\mu}) \label{eq:coll}\,,
\end{align}
where \(\mathrm{d}^3\vec{\pi}_j := (2E_j (2\pi)^{3})^{-1}\mathrm{d}^3 \vec{p}_j\) denotes the Lorentz invariant phase space measure, \(g_L = 1 / 2 + \sin^2(\theta_W)\), \(g_R = \sin^2(\theta_W)\), \(\theta_W\) is the Weinberg angle and the statistical factor \(\phi_{\nu e^{-} \nu e^{-}}\) is given by
\begin{align}
    \phi_{\nu e^{-} \nu e^{-}}(t, p^{\mu},p_{1}^{\mu},p_{2}^{\mu},p_{3}^{\mu}) := \phi_{\nu e^{-} \nu e^{-}}^{+}(t, p^{\mu},p_{1}^{\mu},p_{2}^{\mu},p_{3}^{\mu})
    - \phi_{\nu e^{-} \nu e^{-}}^{-}(t, p^{\mu},p_{1}^{\mu},p_{2}^{\mu},p_{3}^{\mu})
\end{align}
with 
\begin{align*}
    \phi_{\nu e^{-} \nu e^{-}}^{+}(t, p^{\mu},p_{1}^{\mu},p_{2}^{\mu},p_{3}^{\mu}) &:= [1 - f_{e^{-}}(t, p_1)] f_{e^{-}}(t, p_3) \{\mathbb{P}_{e} \varrho(t, p_{2}) \mathbb{P}_{e}, \mathbb{I} - \varrho(t, p)\} \,,\\
    \phi_{\nu e^{-} \nu e^{-}}^{-}(t, p^{\mu},p_{1}^{\mu},p_{2}^{\mu},p_{3}^{\mu}) &:= f_{e^{-}}(t, p_1) [1 - f_{e^{-}}(t, p_3)] \{\mathbb{P}_{e} [\mathbb{I} - \varrho(t, p_{2})] \mathbb{P}_{e}, \varrho(t, p)\}\,.
\end{align*}
Furthermore, \(f_{e^-}\) denotes the electron phase space distribution.
The collision terms are calculated applying the methods described in~\cite{Blaschke:2016xxt} to the current scenario.

In addition to the density matrix for neutrino states, in principle there is also an analogous one for antineutrinos which has to be solved at the same time.
In the following, we assume that the lepton-antilepton asymmetry is of the order of the baryon asymmetry and hence negligible compared to the total phase space densities.
This implies that the antineutrino density matrix behaves the same as the neutrino density matrix and therefore we just have to consider the QKE for neutrinos.

In order to keep track of the temperature \(T_{\gamma}\) of the electromagnetic plasma, we need to solve the continuity equation of the universe
\begin{align}
    \dot{\rho} &= -3H(\rho + P)\,,
\end{align}
where \(\rho\) and \(P\) are the total energy density and total pressure of all radiation species, respectively.
By substituting in the equilibrium expressions for electrons and photons and assuming these particles to be in thermal equilibrium\footnote{This assumption is valid due to the rapid electromagnetic interactions between photons and electrons roughly until the time of last scattering.}, this equation can be reformulated into a differential equation for \(T_{\gamma}\).

\subsection{Numerical Solution of the Quantum Kinetic Equations}
\label{ssec:qke_sol}
In order to prepare the numerical solution of the previously introduced QKE~\eqref{eq:EoM_}, we define a new set of dimensionless variables
\begin{align}
    x(t, p) &:= m_0 a(t) \,, \\
    y(t, p) &:= a(t) p \,,
\end{align}
where we choose \(m_0 = 1\,\mathrm{MeV}\).
Therefore, \(x\) represents the dimensionless scale factor and \(y\) is a dimensionless momentum variable not being redshifted over time, since \(p \propto a^{-1}\).
Moreover, \(x\) takes the role of the reciprocal of the neutrino temperature which is equal to \(T_{\gamma}\) at early times but deviates from it after neutrino decoupling and electron positron annihilation.
Transformed to these new variables eq.~\eqref{eq:EoM_} assumes the form
\begin{align}
    x H \partial_x \tilde{\varrho}(x, y) &= -i [\mathcal{H}(x,y),\tilde{\varrho}(x,y)] + \mathcal{C}[x, y, \tilde{\varrho}(x,y)] \label{eq:EoM}\,,
\end{align}
with \(\tilde{\varrho}\) being the density matrix expressed in the new set of variables.
From now on, we will only refer to this quantity and hence drop the tilde, i.e. \(\tilde{\varrho} \to \varrho\).

In order to integrate eq.~\eqref{eq:EoM}, in principle we had to start at \(x_0 = 0\) and set \(\varrho_{ik}(x_0,y) \equiv 0\).
But since the QKE described in the last section are only valid after the strong phase transition, we have to find a finite starting point \(x_0\) matching all criteria of validity of our equations of motion which are
\begin{enumerate}
    \item Active neutrinos are in thermal equilibrium with the electromagnetic plasma.
    \item Quarks and gluons are bound into hadrons.
    \item Contributions from processes involving muons are negligble.
\end{enumerate}
Furthermore, we assume the sterile neutrino density and correlations between active and sterile neutrinos to be negligible at \(x_0\) such that our initial condition for \(\varrho\) is given by 
\begin{align}
    \varrho(x_0,y) \approx \begin{pmatrix}
        (\exp(y) + 1)^{-1} & 0 \\
        0 & 0 \\
    \end{pmatrix}\,. \label{eq:rho_0}
\end{align}
We found that \(x_0 = 0.01\), i.e. \(T_{\gamma, 0} = 100 \,\mathrm{MeV}\), fulfills these criteria.
For more discussions see appendix~\ref{app:x0}.

We terminate the integration at \(x_1\) which we require to fulfill the following criteria:
\begin{enumerate}
    \item Neutrino interactions are completely frozen out
    \item All free neutrons are bound into light nuclei
    \item The neutrino distribution functions have reached their asymptotic values
    \item The relativistic approximation for the oscillation Hamiltonian and collisions is valid
\end{enumerate}
We found \(x_1 = 50\) to be a suitable final point fulfilling these criteria while still being located in radiation domination.

To solve eq.~\eqref{eq:EoM} in the interval \(X := [x_0, x_1]\), we discretize the momentum space \(\Omega_y = [0,\infty)\) and integrate the resulting set of ordinary differential equations.
To do so, we choose \(N_y\) equidistant points between the minimal and maximal momentum values \(y_{\mathrm{min}}, y_{\mathrm{max}}\) at which we cut off the distribution function.
Choosing a minimal value is necessary because the ultra-relativistic approximation employed within the oscillation Hamiltonian is not valid for all momentum values, especially not for \(y = 0\).
Hence, the minimal \(y\) value is chosen to be \(y_{\mathrm{min}} = 10^{-4}\) in order to still yield a reasonably good approximation for the neutrino energy density.
Furthermore, the maximal momentum value is chosen to be \(y = 20\) since the neutrino distribution at this \(y\) value fulfills
\begin{align}
    f_{\nu}(x, y = 20) &\leq f_{\mathrm{eq}}(x, y = 20) = (\exp(y) + 1)^{-1}\vert_{y = 20} \approx 2 \cdot 10^{-9}\,,
\end{align}
which is sufficiently close to zero.
Therefore, the total, relative error induced within the neutrino energy density needed to calculate our central quantities is of the order
\begin{align*}
    \epsilon_{\mathrm{rel}} := \frac{\vert\rho_{\nu}^{\mathrm{approx}} - \rho_{\nu}\vert}{\rho_{\nu}} \sim 10^{-6}\,.
\end{align*}
The discretized version of \(\Omega_y\) then reads
\begin{align*}
    \tilde{\Omega}_{y}(N_y) := \left\{y_{k} = y_{\mathrm{min}} + k \cdot \Delta y \, \Big\vert \, k \in \{0, \ldots, N_y - 1\}, \Delta y = \frac{y_{\mathrm{max}} - y_{\mathrm{min}}}{N_y - 1} \right\}\,.
\end{align*}
Hence, we arrive at \(N_{y}\) coupled, ordinary, differential equations for the density matrix values at the chosen momentum nodes plus one equation for the photon temperature.
Furthermore, decomposing the Hermitian density matrix into its 4 independent, real components yields a total of \(4 N_{y} + 1\) coupled differential equations which need to be solved\footnote{For the numerical details see appendix~\ref{app:num}}.

\subsection{Calculating the helium Abundance}
\label{ssec:He}
In this section, we present how the \(^4\mathrm{He}\) mass fraction \(Y_{^4\mathrm{He}}\) is estimated from the neutrino distribution functions.
Our explanations and notation closely follow the book~\cite{bernstein_1988} by Bernstein on kinetic theory in an expanding universe.

At first, we define the neutron fraction, i.e.
\begin{align}
    X_{n}(t) := \frac{n_{n}(t)}{n_{n}(t) + n_{p}(t)} \,,
\end{align}
where \(n_{n}\) and \(n_{p}\) are the neutron and proton number densities, respectively.
This choice greatly simplifies the Boltzmann equation for \(n_{n}\) since the \(a^{-3}\) dependence of the number densities cancel.
Furthermore, we find
\begin{align}
    \frac{\mathrm{d}}{\mathrm{d} t} \{a^3(t) (n_{n}(t) + n_{p}(t))\} \equiv \mathrm{const.}\,,
\end{align}
because the baryon number in a comoving volume, \(N_{B} \approx a^3(n_n + n_p)\), is conserved within all relevant processes shortly before neutron freeze out.
These processes are
\begin{align}
    n + e^{+} &\leftrightarrow p + \bar{\nu}_e \\
    n + \nu_{e} &\leftrightarrow p + e^{-} \\
    n &\leftrightarrow p + \bar{\nu}_e + e^{-} \,.
\end{align}
The differential equation governing the evolution of \(X_{n}\) reads~\cite{bernstein_1988}
\begin{align}
    \frac{\mathrm{d}X_{n}(x)}{\mathrm{d}x} = \frac{\lambda_{pn}(x)}{x H} (1 - X_{n}(x)) - \frac{\lambda_{np}(x)}{x H} X_{n}(x) \label{eq:EoM_Xn}\,.
\end{align}
in terms of \(x = m_0 a(t)\).
Here \(\lambda_{np}\) is the thermal interaction rate of all processes converting neutrons to protons, while \(\lambda_{pn}\) is that of all processes converting protons to neutrons.
They are given and discussed in appendix~\ref{app:gamma_n}.

We solve eq.~\eqref{eq:EoM_Xn} from \(x(T = 5\,\mathrm{MeV})\) where neutrons and protons are still in thermal equilibrium up until \(x(T = 0.07\,\mathrm{MeV})\) where deuterium dissociation ceases to be efficient.
During the solution of this differential equation, we interpolate the neutrino distribution functions and temperatures obtained on the grid of \((x,y)\) values using the methods described in the previous subsections.

Finally in order to estimate the produced helium abundance, we convert the neutron fraction into the helium mass fraction, i.e.
\begin{align}
    Y_{^4\mathrm{He}} &= \frac{4 n_{^4\mathrm{He}}}{n_{\mathrm{B}}}
    = \frac{2 n_{n}}{n_{\mathrm{B}}}
    = 2 X_{n} \,,
\end{align}
where we used the assumption that approximately all free neutrons are bound into helium-4 nuclei at the end of BBN.

Of course this method is subject to several approximations especially since we neglect the nuclear reaction rates, hence our estimate cannot be compared directly to observations from~\cite{Aver:2015iza}.
Nevertheless, we can inspect if the different models lead to a relative deviation from the expected value which is on the order of experimental uncertainty or if it exceeds this uncertainty significantly.

\newpage
\section{Predicted Effective Degrees of Freedom and helium Abundance}
\label{sec:results}
Now, we present the results for different benchmark points within the 3 dimensional parameter space.
In the following, we first discuss our results for the pure ADR scenario, the pure scalar field scenario and afterwards for the combination of both effects.
For each chosen benchmark point, we calculate the resulting effective, additional number of degrees of freedom \(\dNeff\) and the estimated helium-4 abundance \(Y_{^{4}\mathrm{He}}\).
These simulated values for \(\dNeff\) are compared to bounds obtained by the Planck collaboration, i.e.
\begin{itemize}
    \item TT + lowE (\(95\%\) CL): \(N_{\mathrm{eff}} = 3.00_{-0.53}^{+0.57}\) \(\Rightarrow\) \(\mathbf{\dNeff \leq 0.57}\),
    \item TT, TE, EE + lowE (\(95\%\) CL): \(N_{\mathrm{eff}} = 2.92_{-0.37}^{+0.36}\) \(\Rightarrow\) \(\mathbf{\dNeff \leq 0.28}\),
    \item TT + lowE + lensing + BAO (\(95\%\) CL): \(N_{\mathrm{eff}} = 3.11_{-0.43}^{+0.44}\) \(\Rightarrow\) \(\mathbf{\dNeff \leq 0.55}\),
    \item TT, TE, EE + lowE + lensing + BAO (\(95\%\) CL): \(N_{\mathrm{eff}} = 2.99_{-0.33}^{+0.34}\) \(\Rightarrow\) \(\mathbf{\dNeff \leq 0.33}\).
\end{itemize}
Here the abbreviations TT, TE, EE, lowE, lensing, BAO refer to different measurement techniques / features of the CMB data (i.e. TT \(\hat{=}\) intensity (temperature) only, TE \(\hat{=}\) temperature + curl free polarization data, EE \(\hat{=}\) curl free polarization data only, lowE \(\hat{=}\) curl free polarization data only at low multipole moments, lensing \(\hat{=}\) grav. lensing measurement, BAO \(\hat{=}\) baryon acoustic oscillations).
Afterwards, we turn towards the helium abundance and compare its deviation for different benchmark points from the expected SM value to the experimental uncertainties on the helium mass fraction from ref.~\cite{Aver:2015iza}, i.e. \(\sigma_{^4\mathrm{He}} = 0.004\).

\subsection{\texorpdfstring{$\dNeff$}{dNeff} and \texorpdfstring{\(Y_{^{4}\mathrm{He}}\)}{YHe-4} in the pure ADR Scenario}
\label{ssec:results_sc_only}
The values of \(\dNeff\) obtained after the full integration of QKEs for different ADR parameters are shown in table~\ref{tab:dNeff_sc_only}.

\begin{figure}
    \centering
    \includegraphics[width=0.7\textwidth]{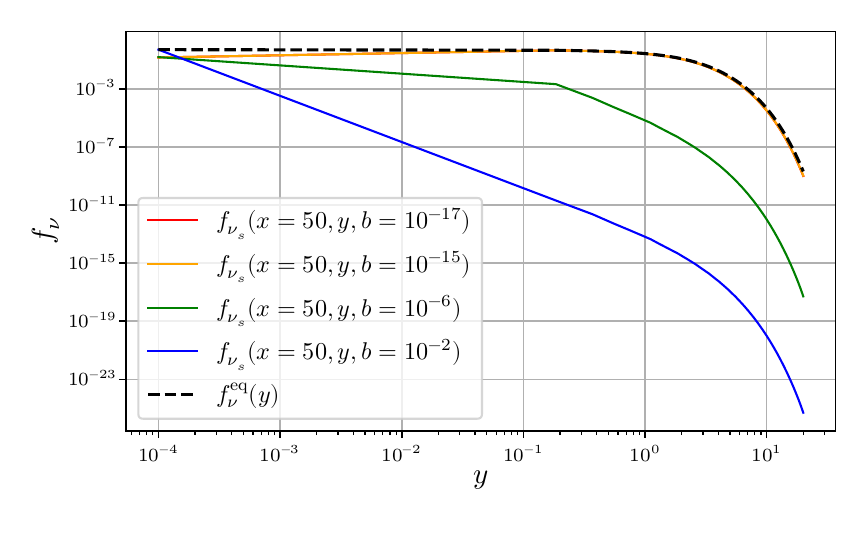}
    \caption{Final (\(x = 50\)) \(\nu_s\) phase space densities (solid) as a function of the comoving momentum \(y\) for various ADR parameters \(b \in \{0, 10^{-17}, 10^{-12}, 10^{-6}\}\) compared to the equilibrium density (dashed).}
    \label{fig:f_nu_s_sc_only}
\end{figure}

\begin{table}
    \centering
    \caption{Estimated additional light degrees of freedom \(\dNeff\) at \(x = 50\) for different ADR parameters \(b\).}
    \label{tab:dNeff_sc_only}
    \begin{tabular}{c|ccccccc}
        \(b\) & \(0\) & \(10^{-17}\) & \(10^{-15}\) & \(10^{-12}\) & \(10^{-6}\) & \(10^{-4}\) & \(10^{-2}\)\\
        \hline
        \(\dNeff\) & \(1.36\) & \(1.36\) & \(1.36\) & \(1.38\) & \(0.04\) & \(0.04\) & \(0.04\)\\
    \end{tabular}
\end{table}
Here, we see that the resulting \(\dNeff\) values for the smallest ADR parameters exceed all Planck bounds by far while bigger \(b\) values lead to excellent agreement with all four bounds \(\dNeff < \{0.28, 0.33, 0.55, 0.57\}\).
Moreover, we can infer that for small \(b\) values the no-ADR scenario is resembled and sterile neutrinos are equilibrated via oscillations.
Turning \(b\) up to much larger values (\(b \gtrsim 10^{-6}\)) leads to a decrease in \(\dNeff\) and sterile neutrinos are not close to equilibrium anymore.
In figure~\ref{fig:f_nu_s_sc_only}, we show the final sterile neutrino distributions compared to the equilibrium distribution for four \(b\) values differing by many orders of magnitude to emphasize this statement.

\begin{figure}
    \centering
    \includegraphics[width=0.7\textwidth]{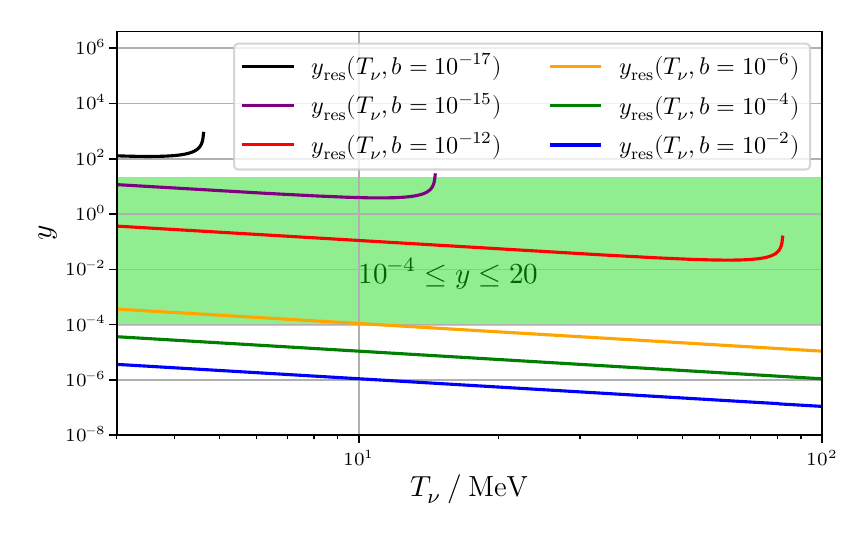}
    \caption{Resonance momentum \(y_{\mathrm{res}}\) plotted against the neutrino temperature \(T_{\nu} \approx T_{\gamma}\) for various ADR parameters \(b \in \{10^{-17}, 10^{-15}, 10^{-12}, 10^{-6}, 10^{-4}, 10^{-2}\}\).
    The relevant momentum region \(10^{-4} \leq y \leq 20\) is shown as the green shaded region.
    All modes above the corresponding resonance curve are subject to strong mixing suppression.}
    \label{fig:y_res}
\end{figure}

The behavior described above can be explained by considering the resonance structure of each parameter configuration.
In figure~\ref{fig:y_res}, we show the resonance momentum \(y_{\mathrm{res}}\) for multiple ADR parameters in the temperature range \(T_{\nu} \in \mathcal{T}_{\nu} := [3, 100] \, \mathrm{MeV}\).
For larger ADR parameters the resonance curve passes through the relevant \(y\) region from above leading to resonantly enhanced \(\nu_e\)-\(\nu_s\) conversion.
Momentum modes located underneath the respective resonance curve neither experience mixing enhancement nor mixing suppression and approximately behave as in the no-ADR scenario.
On the other hand, momentum modes well above the resonance curve, i.e. \(y \gg y_{\mathrm{res}}(T_{\nu})\) \(\forall T_{\nu} \in \mathcal{T}_{\nu}\), are subject to effective mixing suppression since \(\nu_1 \approx \nu_e\) and \(\nu_2 \approx \nu_s\).
Thus, \(\nu_s\) remains unpopulated in this regime.

Note that the strength of this suppression/enhancement effect is momentum dependent since the active as well as the sterile potentials are proportional to \(y\).
Hence, the mixing of neutrinos with small momenta is closer to the vacuum case leading to a faster population of these modes even if the resonance momentum is much smaller.
But since \(\dNeff\) depends more strongly on the high momentum region, this does not affect its estimate significantly.

\begin{figure}
    \centering
    \includegraphics[width = 0.7\textwidth]{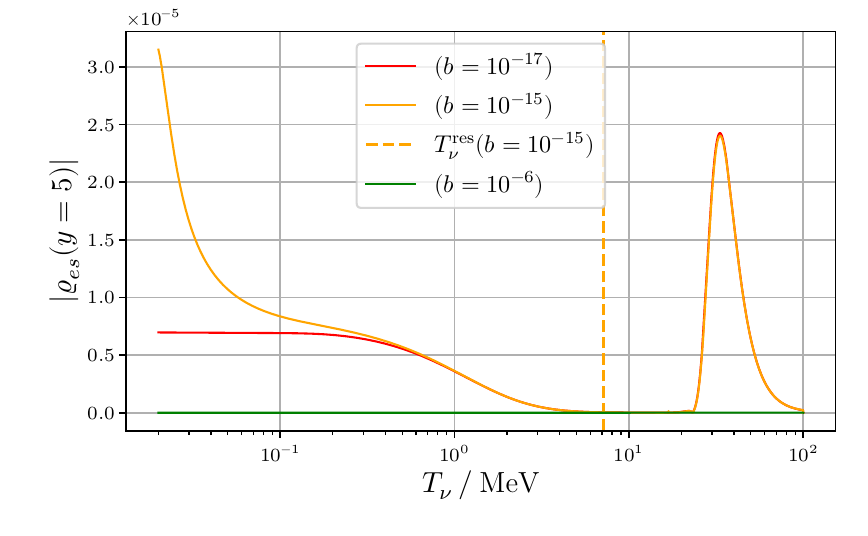}
    \caption{Time evolution of the off-diagonal density matrix element \(\vert \varrho_{es}(y = 5) \vert\) for several ADR parameters.
    The \(y=5\) mode undergoes a resonance at \(T_{\nu} \approx 7\,\mathrm{MeV}\) for \(b=10^{-15}\) leading to a delayed increase of \(\vert \varrho_{es} \vert\) afterwards. The temperature at which this resonance occurs is marked by the orange dashed vertical line.}
    \label{fig:rho_es_sc_only}
\end{figure}

Now one could ask what happens as soon as multiple (high) momentum modes pass through one or two resonances, as is the case for e.g. \(b = 10^{-15}\).
A sterile \(y\) mode can be strongly populated by passing through a resonance, but it is even more important how big the long term mixing around this resonance is.
If the mixing is sufficiently smaller than vacuum mixing, the respective sterile mode experiences some enhancement by the resonance, but it will not reach its equilibrium value. 
On the other hand, if the mixing is very large already before or after passing the resonance, the sterile momentum mode will be equilibrated irregardless of the resonance.
Thus what matters is only the fact whether the relevant momentum modes are subject to mixing suppression for sufficiently long or if they are closer to the (large) vacuum mixing.
This can be seen in figure~\ref{fig:rho_es_sc_only} which shows the temperature evolution of the modulus \(\vert \varrho_{es} \vert := \sqrt{\mathrm{Re}(\varrho_{es})^2 + \mathrm{Im}(\varrho_{es})^2}\) for \(y = 5\).

The off-diagonal element \(\varrho_{es}\) is important since it contains information about the energy transfer from \(\nu_e\) to \(\nu_s\).
Therefore, the plot shows that shortly after \(T = \mathcal{O}(100\,\mathrm{MeV})\) for models with vanishing or small ADR parameters a dip occurs in \(\varrho_{es}\) leading to a significant enhancement of \(\rho_{ss}\), afterwards.
Considering, the curve for \(b = 10^{-15}\), we see that the resonance around \(\mathcal{O}(7\,\mathrm{MeV})\) leads to a significant impact in \(\vert \varrho_{es} \vert\) but doesn't lead to a significant increase of the sterile neutrino density, since it already reached thermal equilibrium.
Moreover, we see that for \(b \geq 10^{-6}\) the off-diagonal matrix elements stay much closer to zero due to sizeable mixing suppression.

Despite the fact that the excess of light degrees of freedom, \(\dNeff\), is a good estimator for the degree of population of sterile neutrinos, it is not sufficient to rely on this number alone.
After neutrino decoupling around \(T_{\gamma} = \mathcal{O}(3\,\mathrm{MeV})\), the active-sterile oscillations could lead to a depletion of the density of active neutrinos which has strong impact on the neutron-proton equilibrium, since fewer \(\nu_e\) lead to an early freeze out of \(n\)-\(p\) reactions and hence to a larger neutron abundance.
This leads to an excess of helium that contradicts the very good agreement of the predictions from standard cosmology and cosmological observations.
Hence, we need to carefully estimate how much helium is produced for our chosen parameter configurations.
To do so, we now estimate the impact of the \(\nu_e\) depletion on nucleosynthesis by proceeding as described in section~\ref{ssec:He} and solve the Boltzmann equation for the neutron fraction \(X_n = n_n / (n_n + n_p)\). 
The neutron fraction \(X_n = n_n / (n_n + n_p)\) can then be translated into the helium mass fraction \(Y_{^4\mathrm{He}} \simeq 2 X_{n}\) at \(T_{\gamma} \approx 0.07\,\mathrm{MeV}\).

The final helium abundances for different ADR parameters are shown and compared to the expected standard value of \(Y_{^4\mathrm{He}}^{\mathrm{std}} \approx 0.227\) in table~\ref{tab:He_sc_only}.
\begin{table}
    \centering
    \caption{Estimated helium abundances for different ADR parameters \(b\) compared to the standard value \(Y_{^4\mathrm{He}}^{\mathrm{SM}} = 0.227 \pm 0.004\).}
    \label{tab:He_sc_only}
    \begin{tabular}{c|cccccccc}
        \(b\) & \(0\) & \(10^{-17}\) & \(10^{-15}\) & \(10^{-12}\) & \(10^{-6}\) & \(10^{-4}\) & \(10^{-2}\)\\
        \hline
        \(Y_{^4\mathrm{He}}\) & \(0.235\) & \(0.235\) & \(0.235\) & \(0.235\) & \(0.227\) & \(0.227\) & \(0.227\)\\
    \end{tabular}
\end{table}
Here, we observe the same consistent picture as for our \(\dNeff\) observable.
Small ADR parameters lead to a deviation \(\Delta Y_{^4\mathrm{He}} = \mathcal{O}(0.01)\) from the SM expectation much larger than the experimental uncertainty \(\sigma_{^4\mathrm{He}} \sim 0.004\) of the observable \(Y_{^4\mathrm{He}}\).
On the other hand, very large ADR parameters, i.e. \(b \gtrsim 10^{-6}\) lead to discrepancies much smaller than \(\sigma_{^4\mathrm{He}}\) and hence would be in agreement with experiment.

\begin{figure}
    \centering
    \includegraphics[width = 0.7\textwidth]{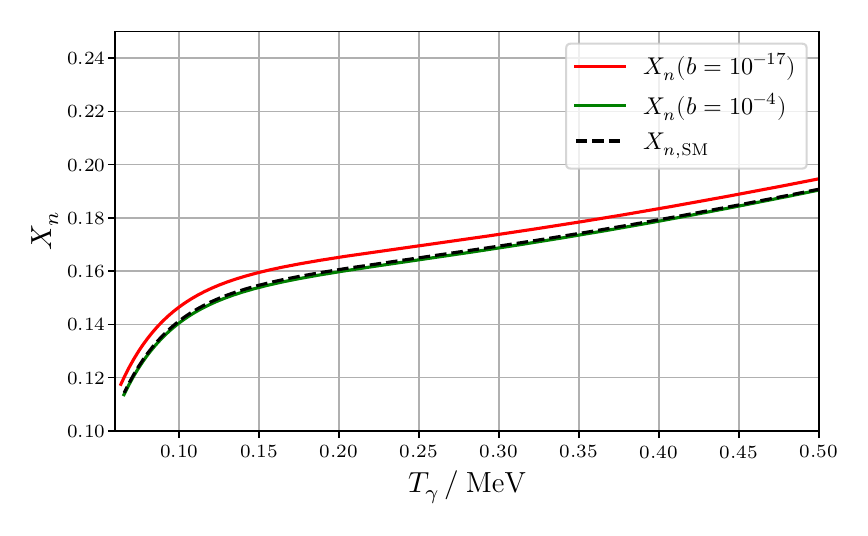}
    \caption{Evolution of the neutron abundance \(X_n\) with respect to the photon temperature \(T_{\gamma}\) for two different ADR parameters, i.e. \(b\in\{10^{-17}, 10^{-4}\}\), after neutron freeze out.
    The dashed black curve represents the SM expectation while the two solid curves correspond to the benchmark scenarios, respectively.}
    \label{fig:X_n_sc_only}
\end{figure}

The argument here is exactly the same as before since the depletion of \(\nu_e\) solely follows from the mixing behavior which in turn is dominantly influenced by the resonance structure.
In figure~\ref{fig:X_n_sc_only}, we compare the temperature evolution of \(X_{n}\) for two different scenarios, i.e. for \(b = 10^{-17}\) and for \(b = 10^{-4}\).
Around the temperature of \(\mathcal{O}(1\,\mathrm{MeV})\) \(X_n\) departs from equilibrium, as expected, in each scenario.
However, \(X_n(b = 10^{-17})\) leaves equilibrium a little earlier and adopts a higher value compared to the SM curve after neutron freeze out.
At the temperature when all neutrons are bound into helium nuclei, this leads to a higher helium abundance.

In the large ADR parameter scenario, i.e. \(b=10^{-4}\), \(X_n\) essentially stays in agreement with the SM curve for the relevant temperatures.
Therefore, the corresponding helium abundance would also be in agreement with the SM expectation within the experimental margin of error.
This is due to the negligible presence of \(\nu_s\) and much higher interaction rates \(\lambda_{np}, \lambda_{pn}\) compared to the previous case caused by the non-dilution of the \(\nu_e\) density.

\subsection{\texorpdfstring{\(\dNeff\)}{dNeff} and \texorpdfstring{\(Y_{^{4}\mathrm{He}}\)}{YHe-4} in the ALP only Scenario}
\label{ssec:results_alp_only}
Next, we look at the behavior of ALP only models, where we turn off the ADR potential and turn on the coupling of the ALP field to the sterile neutrino.
This results in a time dependent, additional mass for the sterile neutrino mass matrix element, \(m_{ss} \rightarrow m_{ss} + m_{s0} \eta(t)\), where \(\eta\) is given by eq.~\eqref{eq:phi_hat}.

We integrate the QKEs for different parameter values shown in table~\ref{tab:dNeff_alp_only}.
The obtained results for \(\dNeff\) show the clear pattern that higher \(m_{s0}\) values and lower \(m_{\phi}\) values are favored by experimental observation.

\begin{table}
    \centering
    \caption{Estimated additional light degrees of freedom \(\dNeff\) at \(x = 50\) for different scalar field parameters.}
    \label{tab:dNeff_alp_only}
    \begin{tabular}{c|ccccccccc}
        $m_{s0}$ / $\mathrm{eV}$ & \multicolumn{3}{l}{$50$} & \multicolumn{3}{l}{$100$} & \multicolumn{3}{l}{$250$} \\
        $m_{\phi}$  / $\mathrm{eV}$ & $10^{-20}$ & $10^{-16}$ & $10^{-12}$ & $10^{-20}$ & $10^{-16}$ & $10^{-12}$ & $10^{-20}$ & $10^{-16}$ & $10^{-12}$ \\
        \hline
        $\Delta N_{\mathrm{eff}}$ & \(1.23\) & \(1.22\) & \(1.30\) & \(0.80\) & \(0.80\) & \(0.93\) & \(0.26\) & \(0.26\) & \(0.30\)
    \end{tabular}
\end{table}

\begin{figure}
    \centering
    \includegraphics[width=0.7\textwidth]{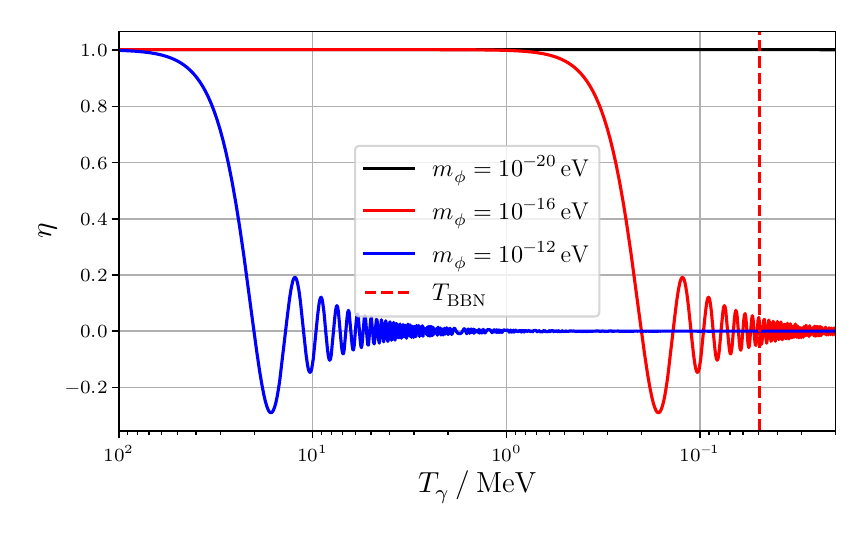}
    \caption{Behavior of the normalized scalar vev \(\eta\) for different ALP mass parameters \(m_{\phi}\) at temperatures in the integration range of the QKEs.}
    \label{fig:phi_hat}
\end{figure}

\begin{figure}
    \centering
    \includegraphics[width=0.7\textwidth]{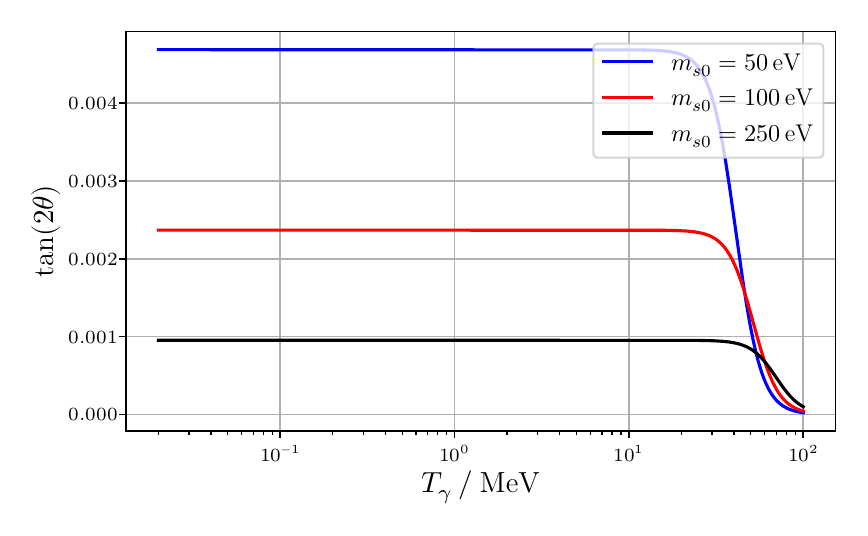}
    \caption{Effective mixing angle between \(\nu_e\) and \(\nu_s\) for different values of \(m_{s0}\) at temperatures in the integration range of the QKEs.
    Here we fixed \(m_{\phi} = 10^{-22}\,\mathrm{eV}\) so that \(\eta\) remains constant for all temperatures of interest.}
    \label{fig:tan2th_alp_only}
\end{figure}

We can explain this by looking at the behavior of the time dependent part \(\propto \eta\) of the sterile mass matrix element which is shown in figure~\ref{fig:phi_hat} and the \(\tan(2\theta)\) of the effective mixing angle shown in figure~\ref{fig:tan2th_alp_only}.
The first plot shows that if the scalar field is too heavy, it starts to oscillate earlier leading to a decrease of the additional mass contribution of the sterile neutrino.
According to the latter figure, smaller sterile masses lead to larger effective mixing angles which in turn lead to faster population of the sterile species.
Therefore, we need a large \(m_{s0}\) and a small \(m_{\phi}\) to reconcile the existence of the sterile species with experimental observations.  

\begin{figure}
    \begin{subfigure}[c]{0.5\textwidth}
        \centering
        \includegraphics[width=\textwidth]{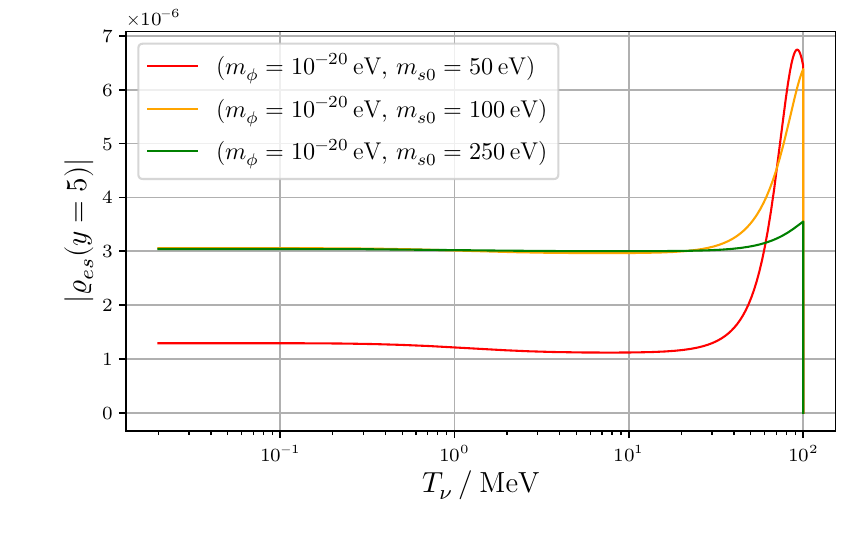}
        \subcaption{\(m_{\phi} = 10^{-20} \, \mathrm{eV}\).}
        \label{fig:rho_es_mphiFix}
    \end{subfigure}
    \begin{subfigure}[c]{0.5\textwidth}
        \centering
        \includegraphics[width=\textwidth]{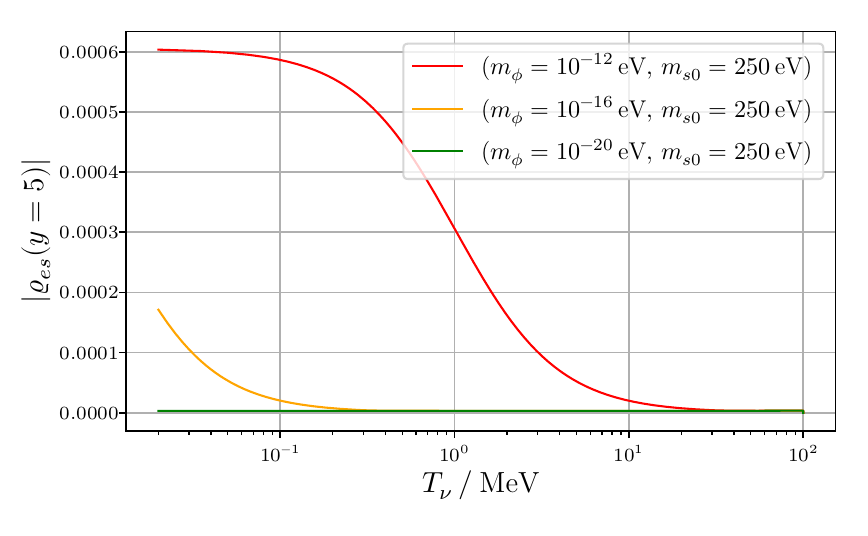}
        \subcaption{\(m_{s0} = 250 \, \mathrm{eV}\).}
        \label{fig:rho_es_ms0Fix}
    \end{subfigure}
    \caption{As figure~\ref{fig:rho_es_sc_only} for the ALP only scenario, for fixed \(m_{\phi}\) (left) and for fixed \(m_{s0}\) (right).
    Shown are three different parameter configurations per panel.}
    \label{fig:rho_es_alp_only}
\end{figure}

These conjectures are supported by figures~\ref{fig:rho_es_mphiFix} and~\ref{fig:rho_es_ms0Fix} showing the off-diagonal element \(\vert \varrho_{es}(y = 5) \vert\).
There we see that after the scalar field has started to oscillate the correlations between active and sterile neutrinos increase, whereas for an overall smaller \(m_{s0}\) parameter the correlations are also overall bigger.

Now, we consider the obtaind helium abundances for the different benchmark points.
In table~\ref{tab:He_alp_only} the resulting values for \(Y_{^4\mathrm{He}}\) are shown.
Here the same pattern arises as for \(\dNeff\) with the difference that our BBN observable is more sensitive to the ALP mass.
For masses \(m_{\phi} \gtrsim 10^{-18} \, \mathrm{eV}\) the condensate oscillates during or before nucleosynthesis has started resulting in a depleted electron neutrino density.
The earlier this oscillation occurs, the more \(f_{\nu_e}\) is depleted and the bigger the deviation of the final helium abundance from the SM value becomes as we can see from table~\ref{tab:He_alp_only}.
Only the benchmark points with \(m_{\phi} \gtrsim 10^{-16} \, \mathrm{eV}\) and \(m_{s0} \gtrsim 100 \, \mathrm{eV}\) are within the uncertainty around the standard value.
\begin{table}
    \centering
    \caption{Estimated helium abundance at \(x = 50\) for different scalar field parameters compared to the standard value \(Y_{^4\mathrm{He}}^{\mathrm{SM}} = 0.227 \pm 0.004\).}
    \label{tab:He_alp_only}
    \begin{tabular}{c|ccccccccc}
        $m_{s0}$ / $\mathrm{eV}$ & \multicolumn{3}{l}{$50$} & \multicolumn{3}{l}{$100$} & \multicolumn{3}{l}{$250$} \\
        $m_{\phi}$  / $\mathrm{eV}$ & $10^{-20}$ & $10^{-16}$ & $10^{-12}$ & $10^{-20}$ & $10^{-16}$ & $10^{-12}$ & $10^{-20}$ & $10^{-16}$ & $10^{-12}$ \\
        \hline
        \(Y_{^4\mathrm{He}}\) & \(0.234\) & \(0.234\) & \(0.235\) & \(0.232\) & \(0.232\) & \(0.233\) & \(0.229\) & \(0.229\) & \(0.229\)
    \end{tabular}
\end{table}

\begin{figure}
    \centering
    \includegraphics[width=0.7\textwidth]{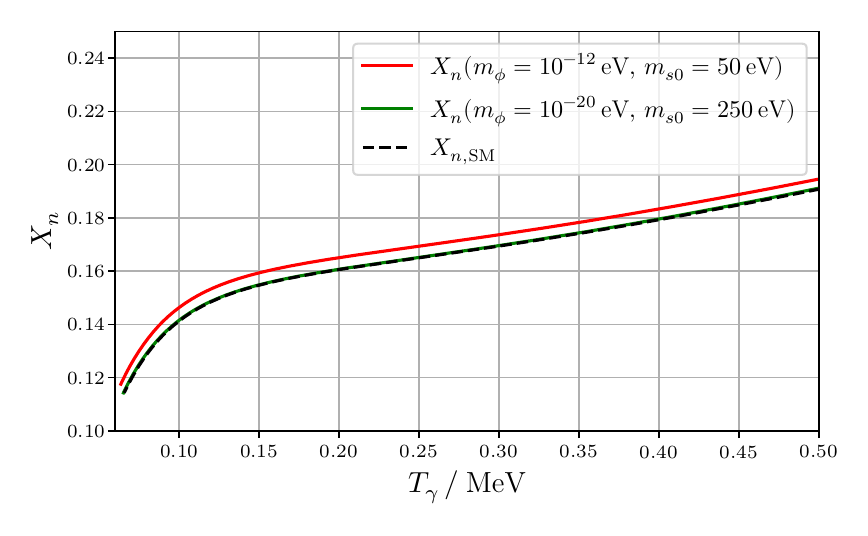}
    \caption{As figure~\ref{fig:X_n_sc_only}, for ALP only scenarios after neutron freeze out.}
    \label{fig:X_n_alp_only}
\end{figure}

To underline this statement we also show the evolution of the neutron fraction for the benchmark points which are least and most compatible with observations in figure~\ref{fig:X_n_alp_only}.
Note that the red curve deviates more significantly from the SM expectation than the green one describing the most compatible parameter configuration under consideration.
The departure of the model curve for the least compatible configuration due to the depletion of electron neutrinos is more prominent than the almost non-existing one in the most compatible case.

As we have seen, a bigger sterile mass matrix element due to the coupling of a scalar field suppresses the mixing of \(\nu_e\) and \(\nu_s\) such that the resulting helium fraction and effective degrees of freedom become compatible with experimental bounds.
Next, we consider the combined ADR and scalar field scenario.

\subsection{\texorpdfstring{\(\dNeff\)}{dNeff} and \texorpdfstring{\(Y_{^{4}\mathrm{He}}\)}{YHe-4} in the combined ADR and ALP Scenario}
\label{ssec:results_sc_alp}
\begin{figure}
    \centering
    \includegraphics[width = 0.7\textwidth]{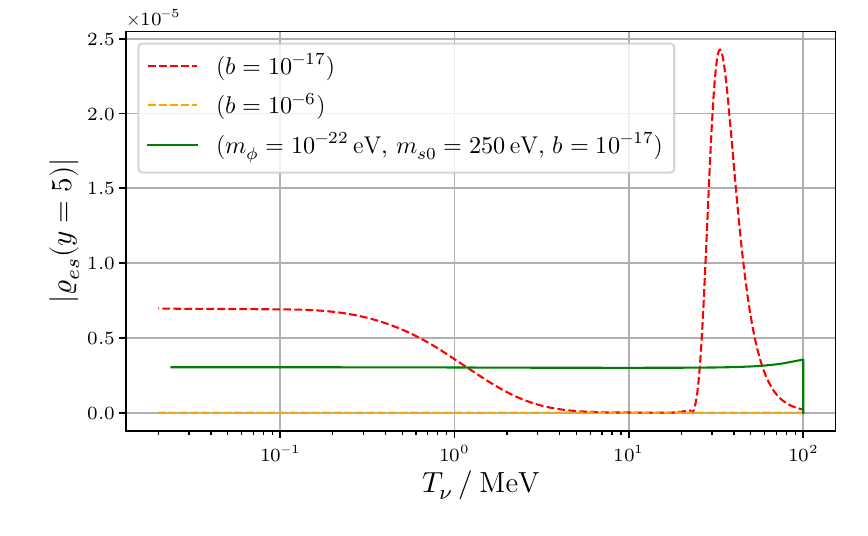}
    \caption{Comparison of the evolution of \(\vert \varrho_{es}(y = 5) \vert\) for the ADR only scenarios \(b \in \{10^{-17}, 10^{-6}\}\) versus the combined scenario with \(m_{\phi} = 10^{-22}\,\mathrm{eV}\), \(m_{s0} = 250\,\mathrm{eV}\) and \(b = 10^{-17}\).
    The ADR only plots are shown as dashed lines, while the line corresponding to the combined scenario is solid.
    Compare Figs.~\ref{fig:rho_es_sc_only} and~\ref{fig:rho_es_alp_only} for the individual ADR and scalar field cases.}
    \label{fig:rho_es_combined}
\end{figure}

In the ADR only case, we have concluded that for sufficiently large ADR parameters \(b\) the equilibration of \(\nu_s\) is suppressed.
Choosing smaller ADR parameters leads to a strong population of \(\nu_s\) and hence large corrections to \(N_{\mathrm{eff}}\) that exceed experimental bounds.
We expect that in the combined scenario even small ADR parameters can be brought into agreement with experiment by invoking the mixing suppression by the scalar field \(\phi\) from section~\ref{ssec:model_alp}.
This expectation is further substantiated from inspecting figure~\ref{fig:rho_es_combined}, which shows a (on average) smaller \(\vert \varrho_{es} \vert\) than for small \(b\) values in the ADR only case.

We confirm this expectation by choosing \(m_{\phi} \sim 10^{-22}\,\mathrm{eV}\) and \(m_{s0} \sim 250\,\mathrm{eV}\).
For this value of \(m_{\phi}\) the scalar field starts oscillating long after nucleosynthesis has ceased and leads to a constant addition to the sterile neutrino mass during the time of integration.
In table~\ref{tab:dNeff_sc_alp}, we display the values of \(\dNeff\) again for \(b \in \{0, 10^{-17}, 10^{-15}, 10^{-12}\}\) with the addition of the scalar field.

\begin{table}
    \centering
    \caption{Estimated additional light degrees of freedom \(\dNeff\) at \(x = 50\) for different ADR parameters \(b\) and \(m_{\phi} \sim 10^{-22}\,\mathrm{eV}\) and \(m_{s0} \sim 250\,\mathrm{eV}\).}
    \label{tab:dNeff_sc_alp}
    \begin{tabular}{c|cccc}
        \(b\) & \(0\) & \(10^{-17}\) & \(10^{-15}\) & \(10^{-12}\)\\
        \hline
        \(\dNeff\) & \(0.25\) & \(0.25\) & \(0.25\) & \(0.26\)\\
    \end{tabular}
\end{table}

\begin{figure}
    \centering
    \includegraphics[width=0.7\textwidth]{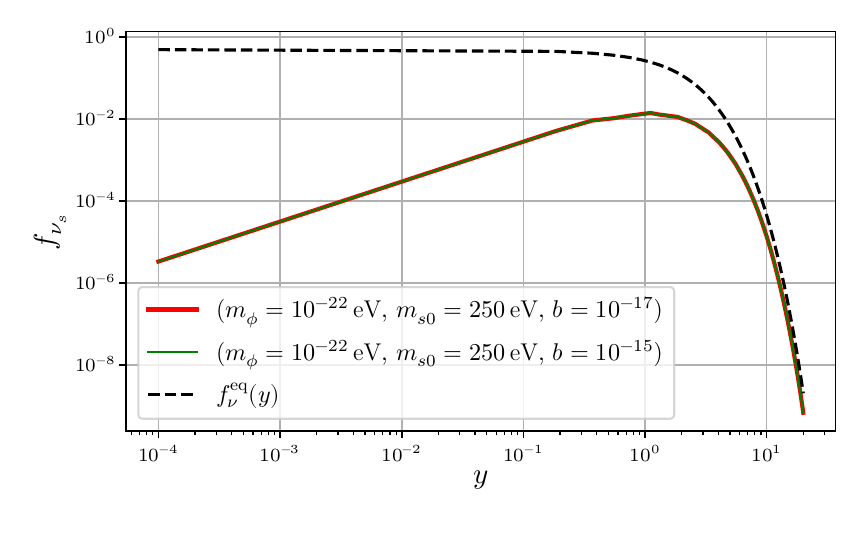}
    \caption{As figure~\ref{fig:f_nu_s_sc_only} for the combined ADR and ALP scenario, for two different ADR parameters.
    The shown \(b\) values are excluded in the ADR only scenario but are reconciled with experiment due to the addition of the scalar field.}
    \label{fig:f_nu_s_combined}
\end{figure}

We see that the values of \(\dNeff\) decrease significantly compared to the pure ADR scenario which is due to mixing suppression because of the additional sterile mass as discussed in the last subsection.
Now the corrections to \(N_{\mathrm{eff}}\) are in agreement with all bounds from Planck, i.e. \(\dNeff < \{0.28, 0.33, 0.55, 0.57\}\).
Thus, even cases where the ADR scenario alone does not explain cosmological observation an additional mass contribution for the sterile neutrino can reconcile them with experiment.
For comparison with the pure ADR case, we show the final neutrino distributions for \(b = 10^{-17}\) and \(b = 10^{-12}\) in figure~\ref{fig:f_nu_s_combined}.

Of course this suppression effect disappears if one allows for higher scalar masses such that the asymptotic sterile mass at \(\mathcal{O}(1\,\mathrm{MeV})\) is reached before BBN or neutrino freeze out or if \(m_{s0}\) is too small in the first place.
Consider for example the benchmark points
\begin{align}
    p_{1} &:= (m_{\phi}, m_{s0}, b) = (10^{-22}\,\mathrm{eV}, \mathbf{50}\,\mathrm{eV}, 10^{-17})\,,\\
    p_{2} &:= (m_{\phi}, m_{s0}, b) = (\mathbf{10^{-12}}\,\mathrm{eV}, 250 \,\mathrm{eV}, 10^{-17})\,, \label{eq:p2}
\end{align}
where we adopted a smaller value for \(m_{s0}\) for \(p_{1}\) and a bigger value for \(m_{\phi}\) for \(p_2\), respectively.
Integrating the QKEs for the first configuration yields
\begin{align}
    \dNeff(p_{1}) &\approx 1.23 \,,
\end{align}
which is slightly smaller than the original result \(\dNeff(b = 10^{-17}) \approx 1.36\) in the pure ADR case, but is still much larger than for \(m_{s0} = 250\,\mathrm{eV}\).
Hence, as expected, a smaller effective sterile mass contribution leads to a reduced mixing suppression.
Increasing the scalar field mass as specified in eq.~\eqref{eq:p2} for the second benchmark point, we get
\begin{align}
    \dNeff(p_2) &\approx 0.30 \,,
\end{align}
which is slightly higher than the value we obtained for \(m_{\phi} = 10^{-22}\,\mathrm{eV}\) but still in agreement with three out of four Planck bounds.
This is because after neutrino freeze out \(\dNeff\) remains constant and \(\phi(t)\) only starts oscillating shortly before.
Thus, we obtain a similar result for \(\dNeff(p_2)\) as for \(m_{\phi} = 10^{-22}\,\mathrm{eV}\).
For \(m_{\phi} \gg 10^{-12} \,\mathrm{eV}\) the scalar field oscillates long enough before neutrino freeze out to imply \(\dNeff(p_2) \rightarrow \dNeff(b)\).

After having discussed the combined scenario for \(\dNeff\), we now turn towards our estimate of the helium abundance.
Here, we expect the same mechanism to apply as in section~\ref{ssec:results_alp_only}:
assuming a scalar field mass of \(m_{\phi} \lesssim 10^{-18}\,\mathrm{eV}\) the oscillatory behavior of the scalar field starts after \(t_{\mathrm{BBN}} \sim 300\,\mathrm{s}\).
Hence, active-sterile oscillations are suppressed depending on the additional mass \(m_{s0}\) during the process of neutron freeze out.

\begin{figure}
    \centering
    \includegraphics[width = 0.7\textwidth]{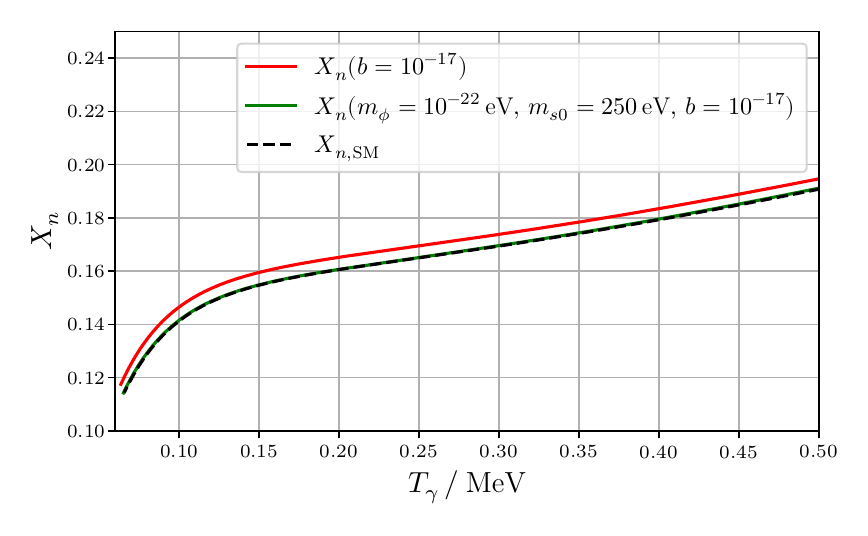}
    \caption{As figure~\ref{fig:X_n_sc_only} comparing the pure ADR case with \(b = 10^{-17}\) with the combined ADR and ALP scenario.}
    \label{fig:X_n_combined}
\end{figure}

We can observe this effect in figure~\ref{fig:X_n_combined} showing the evolution of \(X_n(T)\) for \(m_{\phi} = 10^{-22}\,\mathrm{eV}\), \(m_{s0} = 250\,\mathrm{eV}\) and \(b = 10^{-17}\) compared to the evolution in the pure ADR case with the same \(b\) value.
After adding the scalar field to the pure ADR model, one can no longer distinguish the SM from the model curve within the given margin of error \(\sigma_{^4\mathrm{He}} = 0.004\).
This holds for all chosen ADR parameter configurations with low \(m_{\phi}\) and high \(m_ {s0}\), c.f table~\ref{tab:He_sc_alp}.
\begin{table}
    \centering
    \caption{Estimated helium abundances for different ADR parameters \(b\) and \(m_{\phi} \sim 10^{-22}\,\mathrm{eV}\) and \(m_{s0} \sim 250\,\mathrm{eV}\) compared to the standard value \(Y_{^4\mathrm{He}}^{\mathrm{SM}} = 0.227 \pm 0.004\).}
    \label{tab:He_sc_alp}
    \begin{tabular}{c|cccc}
        \(b\) & \(0\) & \(10^{-17}\) & \(10^{-15}\) & \(10^{-12}\)\\
        \hline
        \(Y_{^4\mathrm{He}}\) & \(0.229\) & \(0.229\) & \(0.229\) & \(0.229\)\\
    \end{tabular}
\end{table}
Hence, in the combined scenario the deviation from the SM value is even smaller than experimental uncertainties \(\Delta Y_{^4\mathrm{He}} \sim 0.002 < \sigma_{^4\mathrm{He}}\).

This effect gets weaker if we choose a lower \(m_{s0}\) or increase the mass of the scalar field up to values of \(m_{\phi} \gg 10^{-18}\,\mathrm{eV}\).
For higher scalar masses, the ALP condensate already oscillates at times before neutron-proton interactions freeze out and hence active-sterile oscillations are not suppressed anymore regardless of the value of \(m_{s0}\).

In order to demonstrate this effect, we again consider the benchmark points \(p_{1}\) and \(p_{2}\) as in the \(\dNeff\) analysis.
For the first point with lower \(m_{s0}\), we again expect a less efficient mixing suppression within the whole integration interval.
This is indeed what we get after integrating the QKEs for the neutrino density matrix and the neutron fraction.
The final helium abundance for \(p_{1}\) amounts to
\begin{align}
    Y_{^4\mathrm{He}}(p_{1}) &\approx 0.234 > 0.229\,,
\end{align}
which is larger than the value for \(m_{s0} = 250\,\mathrm{eV}\) and would be observable in experiments.

For the second benchmark point with lower scalar mass, we again obtain a lower value
\begin{align}
    Y_{^4\mathrm{He}}(p_2) &\approx 0.229\,.
\end{align}
By adopting even smaller scalar mass parameters, we expect this value to approach the pure ADR scenario.
As a consequence, we would end up with more helium-4.

\begin{figure}[!htbp]
    \centering
    \includegraphics[width=0.7\textwidth]{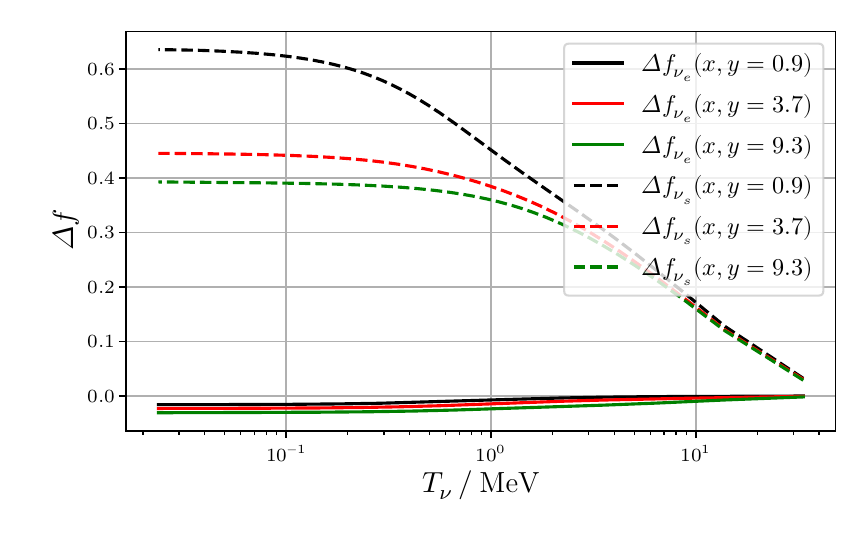}
    \caption{Relative deviations of the \(\nu_e\) and \(\nu_s\) phase space distribution \(\Delta f_{\nu_\alpha}\) at \(p_2\) and \(p_{\mathrm{ref}} = (10^{-22}\,\mathrm{eV},250\,\mathrm{eV}, 10^{-17})\) for different fixed momentum values (black, red, green).
    The distributions are plotted from the onset of scalar field oscillations \(T_{\nu}(t \sim m_{\phi}^{-1})\) down to the final temperature \(T_{\nu,1} = 0.02\,\mathrm{MeV}\), with \(m_{\phi} = 10^{-12}\,\mathrm{eV}\).}
    \label{fig:df_nu_combined}
\end{figure}

This effect can be seen in figure~\ref{fig:df_nu_combined}.
As expected, the \(\nu_e\) density decreases after \(t \sim m_{\phi}^{-1} = 10^{14}\,\mathrm{eV}^{-1}\) while the sterile neutrino density increases, compared to the benchmark point \(p_{\mathrm{ref}}\) with \(m_{\phi} = 10^{-22}\,\mathrm{eV}\) and all other parameters equal to those of \(p_2\).
As a consequence, the higher \(\nu_s\) density leads to an earlier departure of \(X_n\) from equilibrium, while the depleted \(\nu_e\) density causes an even bigger discrepancy between the SM curve and the corresponding model curve due to the smaller \(n-p\) reaction rates.

For higher \(m_{\phi}\) this happens even earlier leading to a more significant increase in \(f_{\nu_{s}}\) (and a corresponding decrease in \(f_{\nu_{e}}\)).

\newpage
\section{Conclusions}
\label{sec:conclusions}
In this paper we have analyzed the impact of altered dispersion relations (ADRs) and couplings to an axion-like scalar field on cosmological bounds for sterile neutrinos.
Both effects have the potential to ameliorate such bounds, depending on the concrete choice of parameters.
In particular, we conclude that ADR parameters in the range needed to give an explanation for \textit{short baseline experiments}, i.e \(b = \mathcal{O}(10^{-17})\), alone are not sufficient to suppress \(\nu_s\) population in the early universe.
We show this by calculating the effective number of additional light degrees of freedom for these parameters and by estimating the amount of helium produced during BBN.
This estimate results in the values
\begin{align}
    \dNeff(b = 10^{-17}) &\approx 1.36\,, \\
    Y_{^4\mathrm{He}}(b = 10^{-17}) &\approx 0.235\,.
\end{align}
Both quantities adopt values higher than allowed by experimental observations, i.e. \(\dNeff \stackrel{\mathrm{Planck}}{\leq} 0.33, 0.57\) and \(Y_{^4\mathrm{He}} = 0.227_{\text{SM pred.}} \pm 0.004\).

In contrast, much larger ADR parameters, like \(b \gtrsim 10^{-6}\), can indeed suppress the \(\nu_s\) population sufficiently to make light sterile neutrinos compatible with early universe cosmology.
For this range of parameters most momentum modes experience mixing suppression because propagation and flavor eigenstates almost coincide with each other resulting in the absence of active sterile oscillations.
Hence, we obtain
\begin{align}
    \dNeff(b \gtrsim 10^{-6}) &< 0.04\,, \\
    Y_{^4\mathrm{He}}(b \gtrsim 10^{-6}) &= 0.227\,.
\end{align}
Considering \(b\)-values differing from the short baseline anomaly scenario may be, for example, motivated by effects making the ADR parameter dependent of the cosmic evolution.

Moreover by adding the influence of an axion-like scalar field \(\phi\) changing the sterile neutrino mass \(m_{ss} \to m_{ss} + m_{s0} \eta(m_{\phi},t)\) via a Yukawa coupling, even very small ADR parameter values can be brought into agreement with experiment.
For a scalar field mass \(m_{\phi} = 10^{-22}\,\mathrm{eV}\) and additional mass amplitude \(m_{s0} = 250\,\mathrm{eV}\), we obtain
\begin{align}
    \dNeff(m_{\phi} = 10^{-22}\,\mathrm{eV}, m_{s0} = 250\,\mathrm{eV}, b = 10^{-17}) &\approx 0.26\,, \\
    Y_{^4\mathrm{He}}(m_{\phi} = 10^{-22}\,\mathrm{eV}, m_{s0} = 250\,\mathrm{eV}, b = 10^{-17}) &\approx 0.229\,,
\end{align}
which is compatible with observations.

While this analysis was carried out for a two neutrino generations framework, i.e. one active and one sterile neutrino, our findings are expected to hold even in scenarios involving greater numbers of generations.
This expectation is justified as it has been found that the active-sterile decoupling can be a generic effect of the model~\cite{Doring:2018cob} at energies higher than the resonance energies.
Also the additional sterile mass from the \(\nu_s\)-\(\phi\) coupling is expected to have the same effect if more neutrino generations are present.
Increasing the diagonal elements of the mass matrix corresponding to the sterile species makes it dominantly diagonal resulting in suppressed active-sterile mixing.

Furthermore, in this analysis we have neglected parameter configurations leading to an early equilibrated species, i.e. at around \(T = \mathcal{O}(100\,\mathrm{MeV})\), for which the integration of the QKEs had to be started much earlier at \(T = \mathcal{O}(1\,\mathrm{GeV})\) as well as finite temperature QED corrections.
The former is justified since the corresponding parameter configurations are not of interest to us since they violate cosmological bounds by definition and would be excluded anyway.
Moreover, due to our findings described in section~\ref{ssec:qke_sol} and~\ref{sec:results}, we don't expect that our conclusions will be different if such an analysis is carried out.

Finite temperature QED corrections are important for precision predictions of the total number of ultra relativistic degrees of freedom~\cite{Bennett:2020zkv} since they can lead to an increase in \(\dNeff\) on the order of magnitude of \(\dNeff = \mathcal{O}(0.1)\).
Here we were solely interested in the impact of light sterile species on the number of additional neutrino generations being mainly influenced by the oscillation Hamiltonian.
Finite temperature QED corrections only have a subleading influence on sterile neutrinos because they do not interact directly with the electromagnetic plasma.

In summary we find that the ADR-only scenario can only explain cosmological observations if one assumes \(b \gtrsim 10^{-6}\). 
The ALP only scenario works well for \(m_{s0} \gtrsim 100 \,\mathrm{eV}\) and \(m_{\phi} \lesssim 10^{-14} \,\mathrm{eV}\), whereas in the combined case also ADR parameters compatible with SBL anomalies (\(b \sim 10^{-17}\)) can be brought into agreement with experimental data for the same \((m_{\phi}, m_{s0})\) configuration as in the ALP only case.

Thus, if sterile neutrinos are discovered at future experiments, ADR effects or a Yukawa coupling to a scalar condensate provide a promising explanation why they did not reach thermal equilibrium in the early universe.
By choosing sufficiently high ADR parameters, high \(m_{s0}\) and low \(m_{\phi}\) these effects lead to a suppression of \(\nu_s\) population regardless of the strength of vacuum mixing between active and sterile neutrinos.

\newpage
\appendix

\section{Evolution of the total Neutrino Density after Decoupling}
\label{app:rho_nu}
To discuss the behavior of the total neutrino energy density after neutrino decoupling, we first prove that the trace of the neutrino density matrix, i.e. the sum of neutrino phase space distributions, is unaffected by neutrino oscillations:
\begin{align}
    (\partial_t - p H \partial_p) \mathrm{Tr}(\varrho(t, p)) &= -i \mathrm{Tr}([\mathcal{H}(t, p), \varrho(t, p)]) + \mathrm{Tr}(\mathcal{C})[t, p, \varrho]\\
    \Leftrightarrow (\partial_t - p H \partial_p) \mathrm{Tr}(\varrho(t, p)) &= \mathrm{Tr}(\mathcal{C})[t, p, \varrho]\,,
\end{align}
where we made use of the fact, that the trace of a commutator always vanishes.
Thus, the sum of neutrino distribution functions is only changed by the expansion of the universe and by neutrino scatterings with the background plasma.
After the freeze out of weak interactions, only the expansion of the universe dilutes the total neutrino density and we get
\begin{align}
    \partial_t \rho_{\nu} = - 4 H \rho_{\nu} = - 3H (\rho_{\nu} + P_{\nu})\,,
\end{align}
where \(\rho_{\nu}\) is the sum of all neutrino energy densities.
Thus, we conclude that for a frozen out neutrino sector the total energy density obeys the cosmic continuity equation, whereas before neutrino freeze out this equation only holds for the total energy density of all radiation species.

\section{\texorpdfstring{Choice of the Initial Scale Factor \(x_0\)}{Choice of the Initial Scale Factor x0}}
\label{app:x0}
In section~\ref{ssec:qke_sol}, we state our choice of the initial dimensionless scale factor to be
\begin{align}
    x_0 = 0.01 \,,
\end{align}
corresponding to an initial neutrino and photon temperature of
\begin{align}
    T_{\nu} = T_{\gamma} = 100\,\mathrm{MeV}\,.
\end{align}
At this temperature, interactions between neutrinos and the eletromagnetic plasma are sufficiently rapid such that \(T_{\nu} = T_{\gamma}\) holds.
Furthermore, we know that the plasma at this temperature is mainly comprised of the following particles
\begin{align}
    e^{-}, e^{+}, \{\nu_{\alpha}\}_{\alpha = e}^{s}, \{\bar{\nu}_{\alpha}\}_{\alpha = e}^{s}, \gamma\,,\label{eq:app:particles}
\end{align}
since the next heavier particles, i.e. muons and pions, are already non relativistic and mainly decayed into these lighter particles.
This dramatically simplifies the collision terms we have to consider for the neutrino Boltzmann equations.
However, in order to verify the validity of the chosen initial time, we have to integrate the Boltzmann equations at much higher temperatures \(T \gg m_{\mu}\) for a set of representative parameter configurations.

In the following, we present an argument why it is still sufficient to consider only the reactions listed in section~\ref{ssec:Boltzmann} for these test runs.
To show that, we decompose the Boltzmann equation for the density matrix
\begin{align}
    xH \frac{\partial \rho}{\partial x}(x,y) &= -i[\mathcal{H}(x,y), \rho(x,y)] + \mathcal{C}[\rho, x, y]\label{eq:app:Boltzmann}\,,
\end{align}
using the \(SU(2)\) generator basis of \(\mathbb{H}(2)\), i.e. the space of Hermitian \(2 \times 2\) matrices.
Thus, we expand all appearing matrices using the extended set of scaled Pauli matrices
\begin{align}
    \tau_0 &= \frac{1}{2} \begin{pmatrix} 1 & 0 \\ 0 & 1 \end{pmatrix}\,, \quad \tau_1 = \frac{1}{2} \begin{pmatrix} 0 & 1 \\ 1 & 0 \end{pmatrix}\,,\\
    \tau_2 &= \frac{1}{2} \begin{pmatrix} 0 & -i \\ i & 0 \end{pmatrix}\,, \quad \tau_3 = \frac{1}{2} \begin{pmatrix} 1 & 0 \\ 0 & -1 \end{pmatrix}\,.
\end{align}
Before applying this decomposition, we rewrite the collision term as the difference of gain and loss terms
\begin{align}
    \mathcal{C}[\rho, x, y] &= \{\Gamma^{+}, \mathbb{I} - \rho\} - \{\Gamma^{-}, \rho\}\,,
\end{align}
where \(\Gamma^{+} \in \mathbb{H}(2)\) is the matrix valued collision rate for gaining a neutrino from the plasma, while \(\Gamma^{-}\in \mathbb{H}(2)\) is the rate for losing a neutrino to the background (or to another momentum state).
Further using the linearity of the anticommutator yields
\begin{align}
    \mathcal{C}[\rho, x, y] &= \{\Gamma^{+}, \mathbb{I}\} - \{\Gamma^{+}, \rho\} - \{\Gamma^{-}, \rho\}\\
    &= 2\Gamma^{+} - \{\Gamma^{+} + \Gamma^{-}, \rho\}\,.
\end{align}
Now, we have to find the components of \(\Gamma^{\pm}\) in our basis.
To achieve this, we note that sterile neutrinos do not interact with the plasma at all which is why \(\Gamma^{\pm}\) takes the form
\begin{align}
    \Gamma^{\pm} = \underbrace{\gamma^{\pm}}_{\geq 0} \mathbb{P}_{a}\,,
\end{align}
where \(\mathbb{P}_{a}\) is the active neutrino projector.
For one active and one sterile neutrino flavor, it is given by
\begin{align}
    \mathbb{P}_{a} = \begin{pmatrix} 1 & 0 \\ 0 & 0 \end{pmatrix} = \tau_0 + \tau_3\,.
\end{align}
This automatically gives us the components of \(\Gamma^{\pm}\)
\begin{align}
    \gamma_{0}^{\pm} = \gamma^{\pm}\,,\quad
    \gamma_{1}^{\pm} = 0\,,\quad
    \gamma_{2}^{\pm} = 0\,,\quad
    \gamma_{3}^{\pm} = \gamma^{\pm}\,.
\end{align}
In the following, we employ the notation
\begin{align}
    \mathbf{a} := (a_0, \vec{a}) := (a_0, a_1, a_2, a_3)\,,
\end{align}
for the vector of components \((a_k)_{k = 0}^{3}\) of a Hermitian matrix \(A\).
Moreover, the components of such a matrix are obtained by taking the scalar product defined on \(\mathbb{H}(2)\) of the matrix itself and the corresponding basis matrix, i.e.
\begin{align}
    a_k = \langle A, \tau_k \rangle := 2\mathrm{Tr}(A \cdot \tau_k)\,.
\end{align}
The density matrix and Hamiltonian components are denoted as follows
\begin{align}
    \rho &= \sum_{k = 0}^{3} \varrho_k \tau_k\,,\\
    \mathcal{H} &= \sum_{k = 0}^{3} h_k \tau_k\,.
\end{align}
Substituting this decomposition into eq.~\eqref{eq:app:Boltzmann} yields the 4 coupled equations
\begin{align}
    xH \frac{\partial \varrho_0}{\partial x} &= 2\gamma^{+} - (\gamma^{+} + \gamma^{-})(\varrho_0 + \varrho_3)\\
    xH \frac{\partial \vec{\varrho}}{\partial x} &= \vec{h} \times \vec{\varrho} - (\gamma^{+} + \gamma^{-})\vec{\varrho} + (2\gamma^{+} - (\gamma^{+} + \gamma^{-})\varrho_0) \vec{e}_3\,,
\end{align}
with \(\vec{e}_3 = (0,0,1)^T\).

Now, we are able to argue that taking into account less reactions into our collison terms yields an upper limit for the starting temperature\footnote{An upper limit on \(T_{\nu}\) corresponds to a lower limit for the starting scale factor \(x_0\).}.
This becomes appearant by considering the evolution equations for the real and imaginary parts of the off-diagonal density matrix element \(\varrho_1, \varrho_2\) as well as of the sterile neutrino component \(\rho_{ss} = (\varrho_0 - \varrho_3) / 2\)
\begin{align}
    xH \frac{\partial \varrho_1}{\partial x} &= (\vec{h} \times \vec{\varrho})_1 - (\gamma^{+} + \gamma^{-})\varrho_1\,,\\
    xH \frac{\partial \varrho_2}{\partial x} &= (\vec{h} \times \vec{\varrho})_2 - (\gamma^{+} + \gamma^{-})\varrho_2\,,\\
    xH \frac{\partial \rho_{ss}}{\partial x} &= -(\vec{h} \times \vec{\varrho})_3 \,.
\end{align}
At \(T \gg 100 \,\mathrm{MeV}\), the total neutrino interaction rate \(\gamma^{+} + \gamma^{-}\) becomes sufficiently large leading to an exponential dampening of the off-diagonal matrix elements.
Consequently, the sterile neutrino distribution remains constant at these times since \(\varrho_{1,2} \approx 0\) implies \(\partial \rho_{ss} / \partial x \approx 0\).
Thus, if we start with an unpopulated sterile species at high temperatures it will only start being populated as soon as the interaction rate of \(\nu_e\) becomes weak enough such that oscillations aren't suppressed anymore.

If we neglect some of the processes contributing to \(\gamma^{\pm}\) the temperature where \(\gamma^{+} + \gamma^{-}\) is big enough to suppress oscillations gets shifted to higher values, i.e. smaller \(x\).
Therefore, it is safe to estimate a suitable starting temperature (scale factor) using an incomplete set of interactions since the actual proper estimate will be lower (higher).
Of course, in case this procedure yields a starting temperature \(T_{\nu} > 100\,\mathrm{MeV}\), we need to include the full set of reactions for the actual integration process or deliver good arguments why \(x_0 = 0.01\) is still a good choice.
In the following, we present the results of our test runs.

Solving the set of Boltzmann equations in the high temperature regime is much simpler because active neutrinos are in thermal equilibrium with the electromagnetic plasma.
Therefore, we use
\begin{align}
    T_{\nu} = T_{\gamma} = \frac{\mathrm{MeV}}{x}\,,\\
    f_{\nu_e}(x, y) = f_{\mathrm{eq}}(y)\,,
\end{align}
to precalculate and interpolate the collision terms.
Furthermore, we don't have to solve the evolution equation for the photon temperature.

\begin{figure}
    \begin{subfigure}[c]{0.5\textwidth}
        \centering
        \includegraphics[width=\textwidth]{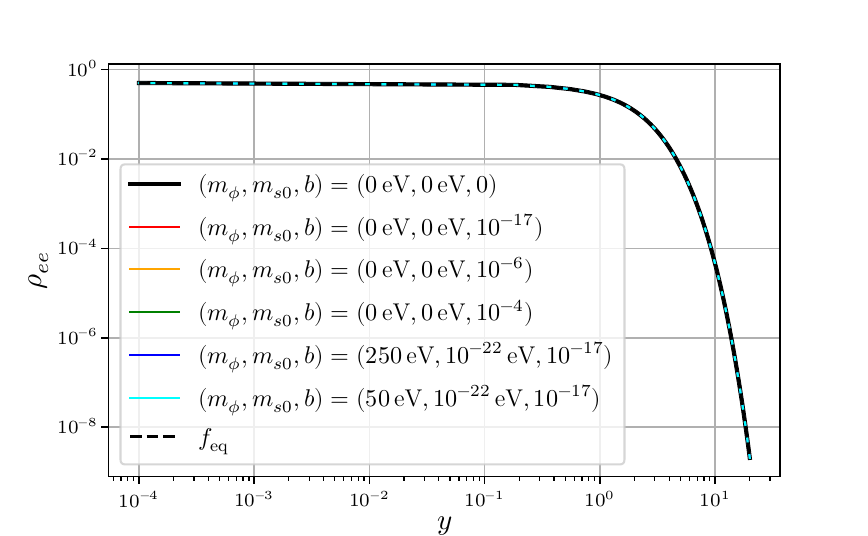}
        \subcaption{\(\rho_{ee}\)}
        \label{fig:app:rho_nu_ee}
    \end{subfigure}
    \begin{subfigure}[c]{0.5\textwidth}
        \centering
        \includegraphics[width=\textwidth]{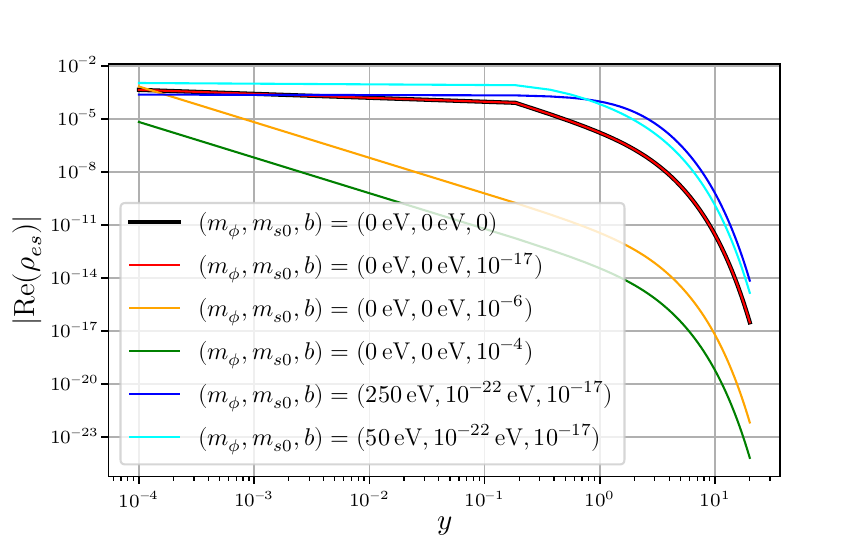}
        \subcaption{\(\mathrm{Re}(\rho_{es})\)}
        \label{fig:app:rho_nu_es1}
    \end{subfigure}
    \begin{subfigure}[c]{0.5\textwidth}
        \centering
        \includegraphics[width=\textwidth]{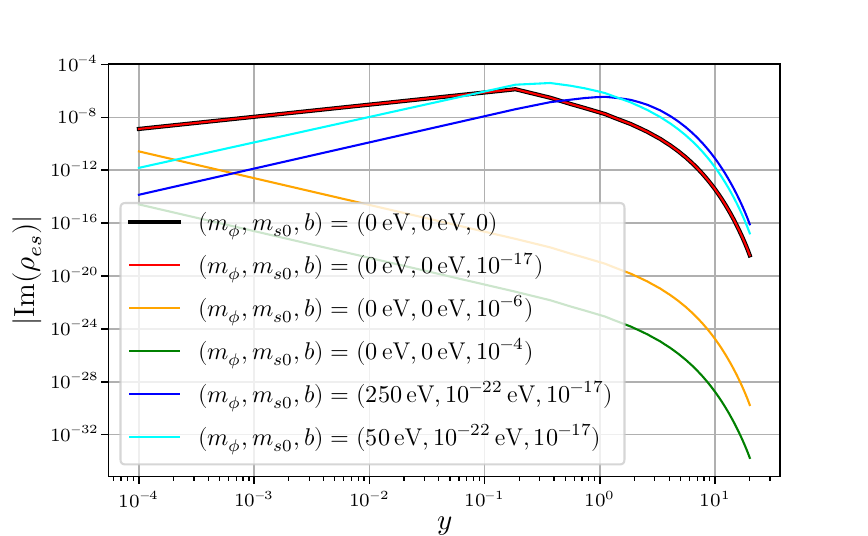}
        \subcaption{\(\mathrm{Im}(\rho_{es})\)}
        \label{fig:app:rho_nu_es2}
    \end{subfigure}
    \begin{subfigure}[c]{0.5\textwidth}
        \centering
        \includegraphics[width=\textwidth]{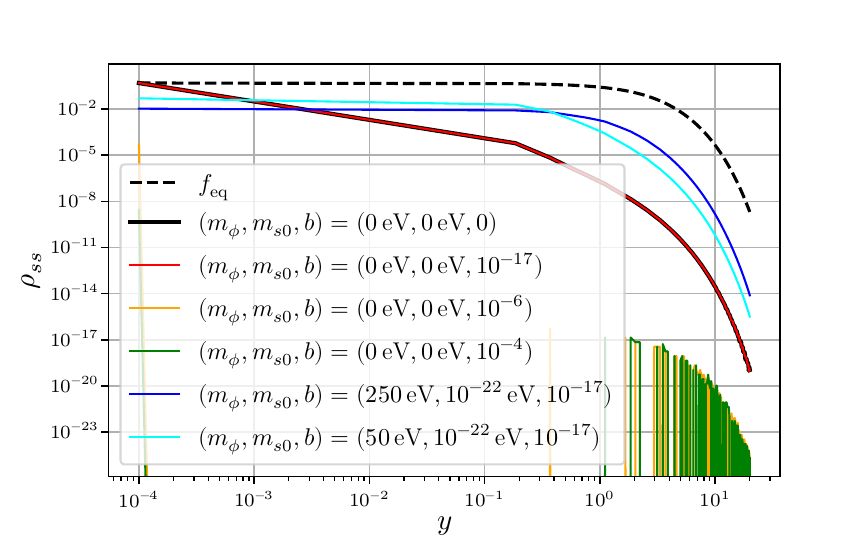}
        \subcaption{\(\rho_{ss}\)}
        \label{fig:app:rho_nu_ss}
    \end{subfigure}
    \caption{Density matrix components at \(x_0 = 0.01\) plotted against the dimensionless momentum \(y\) for different models displayed in the legends.}
    \label{fig:app:rho_nu}
\end{figure}

\begin{figure}
    \centering
    \includegraphics[width=0.7\textwidth]{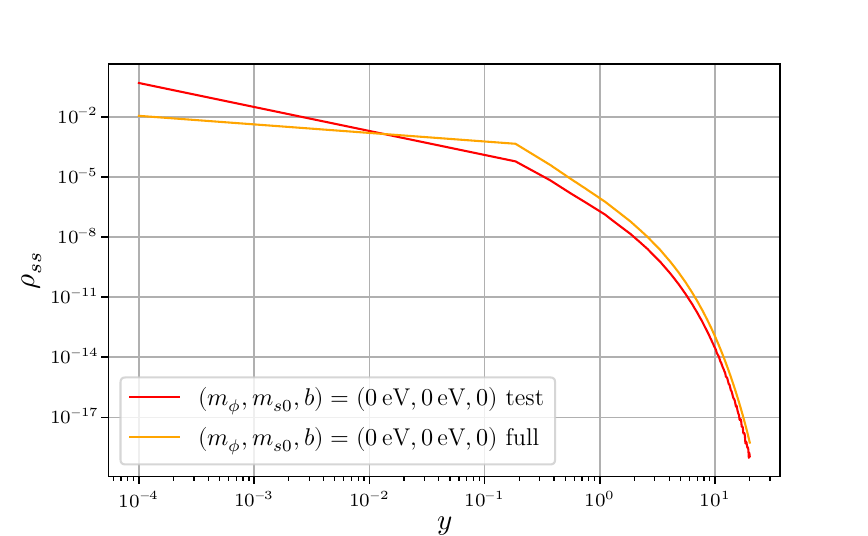}
    \caption{Comparison of the sterile neutrino densities obtained from the test run at \(x_0 = 0.01\) (red) and from the full run after 2 steps at \(x = x_0 + 10^{-6}\) (orange).}
    \label{fig:app:comp}
\end{figure}

In order to ensure that \(\gamma^{+} + \gamma^{-}\) is large enough such that oscillations are suppressed, we choose \(T = 10\,\mathrm{GeV}\) as our initial temperature and integrate until \(T = 100\,\mathrm{MeV}\).
To simplify the procedure, we use the findings from above and only consider reactions with the particles given in~\eqref{eq:app:particles}.\\
In figure~\ref{fig:app:rho_nu}, we show all components of the density matrix at \(x_0 = 0.01\) for a set of different models.
Inspecting figure~\ref{fig:app:rho_nu_ee} shows that the electron neutrino phasespace distribution perfectly aligns with its equilibrium value as expected.
Furthermore, we observe that for models with high ADR parameters the sterile neutrino distribution is very close to 0 (cf.~\ref{fig:app:rho_nu_ss}) such that numerical oscillations below machine precision take over.
For much smaller ADR parameters \(\rho_{ss}\) can get very close to equilibrium especially for small momenta already at \(x_0 = 0.01\) indicating that \(\rho_{ss}(x_0) \equiv 0\) isn't valid.
Turning on the effective sterile mass contribution from the coupling to the scalar field fixes this issue and \(\rho_{ss}\) is close enough to 0 for our initial assumption to hold.
The same conclusions can be drawn from~\ref{fig:app:rho_nu_es1} and~\ref{fig:app:rho_nu_es2} showing the off-diagonal density matrix elements which quantify the correlations between \(\nu_e\) and \(\nu_s\).
In figure~\ref{fig:app:comp}, we compare the final sterile neutrino distribution from the test run at \(x_0 = 0.01\) to the distribution obtained in the full run at \(x = 0.010001\) using the assumption that the sterile neutrino density vanishes at \(x_0\).
Here we can see that the sterile density already approaches and even surpasses the values obtained in the test run at some momenta, hence we conclude that the error introduced by using the \textit{wrong} initial value is small.

In conclusion, we see that for sufficiently high ADR parameters or effective sterile masses the initial condition
\begin{align}
    \rho(x_0 = 0.01, y) \approx \begin{pmatrix}
        f_{\mathrm{eq}}(y) & 0 \\ 0 & 0
    \end{pmatrix}\,,\label{eq:app:dm0}
\end{align}
is a good approximation.

Only for situations where \(b\) is too small or mixing suppression through the additional sterile mass is not sufficient, we obtain significant sterile neutrino densities already at \(x_0 = 0.01\).
Thus, in order to achieve a reliable estimate of the actual sterile neutrino density we need to start the integration much earlier.
However, since these models are already excluded for the starting point \(x_0\) (cf.~\ref{sec:results}) choosing an earlier \(x_0\) would even lead to higher or equal sterile neutrino densities.
This is because oscillations lead to an equilibration of the \(\nu_e\) and \(\nu_s\) number, while collisions are pumping new electron neutrinos into the system as long as their number deviates from the thermal equlibrium values.
Therefore, we consider the obtained sterile neutrino distributions as lower bounds of the actual values.
Moreover, it can be seen by comparing the density matrix components from the test run at \(x_0\) to the components obtained in the full run after a few steps that they almost coincide.
Hence, the estimate of \(\rho(x,y)\) obtained using \(x_0 = 0.01\) for the problematic models is adequate for our purposes.

\section{Numerical Methodology}
\label{app:num}
We solve this set of differential equations using \texttt{C++} and the gnu scientific library~\cite{gsl} (GSL).
In detail, a fixed step size \texttt{Burlisch-Stoer} integrator is employed and we take \(\mathcal{O}(10^4)\) steps between the initial (\(x_0 = 0.01\)) and final \textit{time} (\(x_1 = 50\)).

Moreover, the calculation of all numerical integrals is carried out using the closed Newton-Cotes formula of order 6, which is why we choose \(N_{y} = 109\).
If the density matrix is evaluated at momenta between the nodes defined by \(\tilde{\Omega}_y\), we interpolate it with a cubic Steffen Spline\footnote{The Steffen Interpolation Method guaranties monotonicity of the cubic function between two neighboring data points.}~\cite{steffen_spline}.

\section{Thermal Interactionrates of Neutron-Proton Processes}
\label{app:gamma_n}
The thermal reaction rates needed for the solution of the Boltzmann equation for the neutron abundance are given by~\cite{Weinberg:1972kfs}
\begin{align}
    \lambda_{np}(x) &= \lambda_{n + e^{+} \rightarrow p + \bar{\nu}_{e}} + \lambda_{n + \nu_e \rightarrow p + e^{-}} + \lambda_{n \rightarrow p + \bar{\nu}_e + e^{-}}\,,\\
    \lambda_{pn}(x) &= \lambda_{n + e^{+} \leftarrow p + \bar{\nu}_{e}} + \lambda_{n + \nu_e \leftarrow p + e^{-}} + \lambda_{n \leftarrow p + \bar{\nu}_e + e^{-}}\,,
\end{align}
with the partial rates
\begin{align}
    \lambda_{n + \nu_e \rightarrow p + e^{-}}(x) &= A \int\limits_{0}^{\infty} p_{\nu}^2 p_{e} E_{e} [1 - f_{e}(E_e, T(x))] f_{\nu_e}(x, p_{\nu} a(x)) \,\mathrm{d}p_{\nu}\,, \label{eq:l_np_1}\\
    \lambda_{n + e^{+} \rightarrow p + \bar{\nu}_{e}}(x) &= A \int\limits_{0}^{\infty} p_{e}^2 p_{\nu} E_{\nu} [1 - f_{\nu_e}(x, p_{\nu} a(x))] f_{e}(E_e, T(x)) \,\mathrm{d}p_{e}\,,\label{eq:l_np_2}\\
    \lambda_{n \rightarrow p + e^{-} + \bar{\nu}_e}(x) &= A \int\limits_{0}^{p_0} p_{e}^2 p_{\nu} E_{\nu} (1 - f_{\nu_e}(x, p_{\nu} a(x))) [1 - f_{e}(E_e, T(x))] \,\mathrm{d}p_{e}\label{eq:l_np_3}\,,
\end{align}
where \(p_{\nu}\) is the electron neutrino momentum, \(p_e\) and \(E_e\) are the electron momentum and energy, respectively, and the common constant \(A\) is determined via the mean lifetime of the neutron in its rest frame \(\tau_n\), i.e.
\begin{align}
    A \approx \left\{0.0157 \cdot \tau_n (\Delta m)^5\right\}^{-1}\,, \quad \mathrm{with} \quad \Delta m = m_n - m_p\,.
\end{align}
The interaction rates for the back reactions are obtained from eqs.~\eqref{eq:l_np_1} to~\eqref{eq:l_np_3} by replacing \(f \rightarrow 1 - f\) for all phase space distributrion functions.
Furthermore, the upper integration bound in eq.~\eqref{eq:l_np_3} is given by \(p_0 = \sqrt{\Delta m^2 - m_e^2}\) and the following energy conservation relations are imposed:
\begin{enumerate}
    \item eq.~\eqref{eq:l_np_1}: \(E_e = E_{\nu} + \Delta m\,,\)
    \item eq.~\eqref{eq:l_np_2}: \(E_{\nu} = E_e + \Delta m\,,\)
    \item eq.~\eqref{eq:l_np_3}: \(E_{\nu} = \Delta m - E_e\,.\)
\end{enumerate}


\end{document}